\documentclass[a4paper,fleqn,usenatbib]{mnras}\parskip=2.0mm

\usepackage{amsmath}    % Advanced maths commands
\usepackage{txfonts}
\usepackage[T1]{fontenc}
\usepackage{ae,aecompl}

\usepackage{lipsum,graphicx,multicol}
\usepackage{multirow}

\usepackage[justification=centering]{caption}
\usepackage{graphicx}    % Including figure files

\usepackage{amssymb}    % Extra maths symbols
\usepackage{tabulary}
\title[Probing possible effects of sharing CGMs in pairs]{Probing possible effects of circumgalactic media on the metal content of galaxies through the mass-metallicity relationship} % mass-metallicity 格式

\author[Sai Zhai et al.]{
	Sai Zhai,$^{1,2}$
	Yong Shi,$^{1,2}$\thanks{Email:yong@nju.edu.cn}
	Jianhang Chen,$^{1,3}$
	Longji Bing,$^{1,4}$
	Yanmei Chen,$^{1,2}$\\
	\newauthor
	Xiaoling Yu,$^{1,2}$
	Songlin Li,$^{1,2}$
	\\
	$^{1}$School of Astronomy and Space Science, Nanjing University, Nanjing 210093, China\\
	$^{2}$Key Laboratory of Modern Astronomy and Astrophysics (Nanjing University), Ministry of Education, Nanjing 210093, China\\
	$^{3}$European Southern Observatory, Karl-Schwarzschild-Strasse 2, D-85748 Garching bei Muenchen, Germany\\
	$^{4}$Aix Marseille Université, CNRS, CNES, LAM (Laboratoire d’Astrophysique de Marseille), 13013, Marseille, France\\
}

\date{Accepted XXX. Received YYY; in original form ZZZ}

\pubyear{2019}

\begin{document}
	\label{firstpage}
	\pagerange{\pageref{firstpage}--\pageref{lastpage}}
	\maketitle
	
	\begin{abstract}\parskip=2.0mm

	The circumgalactic medium (CGM) connects the gas between the interstellar medium (ISM) and the intergalactic medium, which plays an important role in galaxy evolution. We use the stellar mass-metallicity relationship to investigate whether sharing the CGM will affect the distribution of metals in galaxy pairs. The optical emission lines from the Sloan Digital Sky Survey Data Release (SDSS DR7) are used to measure the gas-phase metallicity. We find that there is no significant difference in the distribution of the metallicity difference between two members in star forming-star forming pairs ($\rm \Delta log(O/H)_{diff}$), metallicity offset from the best-fitted stellar mass-metallicity relationship of galaxies in pairs ($\rm \Delta log(O/H)_{MS}$), as compared to ``fake'' pairs. By looking at $\rm \Delta log(O/H)_{diff}$ and $\rm \Delta log(O/H)_{MS}$ as a function of the star formation rate (SFR), specific star formation rate (sSFR), and stellar mass ratio, no difference is seen between galaxies in pairs and control galaxies. From our results, the share of the CGM may not play an important role in shaping the evolution of metal contents of galaxies. 
		
	\end{abstract}
	
	
	%%%%%%%%%%%%%%%%%%%%%%%%%%%%%%%%%%%%%%%%%%%%%%%%%%
	
	%%%%%%%%%%%%%%%%% BODY OF PAPER %%%%%%%%%%%%%%%%%%
	
	\section{Introduction}
	
	    Galaxies are surrounded by the multi-phase gas: the circumgalactic medium (CGM), which extends beyond the interstellar medium (ISM) but within the virial radius \citep{2017ARA&A..55..389T}. Galactic scale outflow is common in the active star forming galaxies \citep{ 2009ApJ...692..187W, 2010ApJ...717..289S, 2010AJ....140..445C}, which drives the metals produced in the star-forming region to the CGM \citep{1990ApJS...74..833H, 2011Sci...334..948T}. As a result, even though the CGM is diffuse and thin, it contains the same, or even more amount of baryons than the galaxy disk in spirals \citep{2011Sci...334..948T,10.1093/mnras/staa358, 2014ApJ...792....8W}. Besides outflows, the galactic inflow/accretion of CGMs deposit fresh gas fuel for star formation in disks. This recycling of the material between CGM and disks through outflows and inflows/accretion is one of the key physical processes that shape galaxy formation and evolution \citep{2017ARA&A..55..389T}. In this study, we employ the mass-metallicity relationship to probe the possible effects of such recycling on galaxy evolution. 

	The correlation between the stellar mass-metallicity (hereafter MZR) is one of the most important relationships to study the formation and evolution of galaxies \citep{2006ApJ...644..813E,2013ApJ...779..102K}. Stellar mass estimates the amount of gas that has turned into stars. Metal traces the star formation history \citep{2011ApJ...730..137Z}. The relationship between them can be used to study the physical processes related to the formation and distribution of metals.	
	
	    The analysis of MZR starts from irregular, blue compact galaxies \citep{1979A&A....80..155L}. The work by \cite{2004ApJ...613..898T} has used 53,000 star-forming galaxies to find a nice relationship between stellar mass and metallicity. The trend below $10^{10.5} \ M_{\odot}$ is steep, and above $10^{10.5} \ M_{\odot}$ becomes flat. They pointed out that galactic wind takes an important role in shaping the trend of the MZR. Over decades, studies of MZR have extended to high-redshift \citep{2014MNRAS.437.3647Y,2014A&A...563A..58T,2016ApJ...828...67L,2016ApJ...822...42O,2018ApJ...858...99S}, dwarf galaxies  \citep{2006ApJ...647..970L,2019ApJ...877....6B} and simulation \citep{2007ApJ...655L..17B, 2016MNRAS.456.2140M}. \cite{2018MNRAS.481.1690C} presents that in the local universe, outflow causes the loss of 78\% metal. However, the origin of the stellar mass-metallicity relationship is still uncertain. At least, four main physical processes affect the MZR: the low star formation efficiency of low stellar mass galaxies \citep{2009A&A...504..373C}, the infall and outflow of unenriched/enriched gas \citep{2008MNRAS.385.2181F, 2002ApJ...581.1019G}, and variation of initial mass function (IMF) under different physical conditions \citep{2007MNRAS.375..673K,8224587}.
	
	Compared to isolated galaxies, galaxies in pairs may share a common CGM so that their evolution may be linked to some extent for the following reasons. First, galactic scale activities, such as outflow, inflow, accretion, and recycling occur in the region of the CGM, which regulates the distribution of gas and metal \citep{1990ApJS...74..833H, 2011Sci...334..948T, 2017ARA&A..55..389T}. Second, the CGM contains the majority of baryon in the galaxy, which might provide material for star formation \citep{2014ApJ...792....8W}. Third, for galaxies in pairs that also interact with each other, they already share the CGM to a degree and the interaction between them helps to regulate the distribution of gas and metal, which enhances the potential link. For example, \cite{2018ApJ...868..132P} finds that the interaction between pairs enhances the star formation rate (SFR). Satellite galaxies located in dense environments tend to hold more metals because the metal-rich gas from the high-density region enhances the metallicity of satellites \citep{2014MNRAS.438..262P}. In \cite{2006AJ....131.2004K}, they show that the merging process triggers the infall of metal-poor gas, thus decreasing the metallicity in the central regions of galaxies.

	 In this paper, we search for the metallicity difference between galaxy pairs, which may shed light on the roles of CGMs in driving galaxy formation and evolution. To differentiate the effect of galaxy interaction from that of sharing CGM in affecting the metals of galaxies, we expand the definition of pairs to be those with separations < 300 kpc and choose 150 kpc \citep{Bustamante2020} as the demarcation point to determine whether there is an interaction between pairs or not. In addition, besides star forming-star forming pairs (SF-SF pairs), we add another type of galaxy pairs that contains one star forming member galaxy and one passive member galaxy (SF-Passive pairs), which has no CGM sharing. This is because passive galaxies contain less ionized gas in the CGM than star forming galaxies \citep{2011Sci...334..948T}.
 Two important parameters are included to quantitatively measure the metallicity difference between pairs. One is the metallicity difference between two members in pairs ($\rm \Delta log(O/H)_{diff}$, in equation \ref{metallicity difference}), another is the metallicity offset from the best-fitted stellar mass-metallicity relationship of galaxies in pairs ($\rm \Delta log(O/H)_{MS}$, in equation \ref{metallicity offset}). We describe the observation and data reduction in \S2. In \S3, we show the MZR and metal difference of pairs from SDSS DR7. In \S4, we discuss the incompleteness of the data and the physical mechanism. The conclusion is in \S5. 
	We adopt the flat $\rm \Lambda CDM$ model, $h=0.677, \Omega_{m}=0.307$ \citep{2016A&A...594A..13P}.

	\section{Observations and Data Reduction}
	
	 The Sloan Digital Sky Survey (SDSS, \cite{2000AJ....120.1579Y}) uses the 2.5-m optical telescope located at Apache Point Observatory (APO, \cite{2006AJ....131.2332G}), began in 2000, which aims to obtain the multi-spectral imaging, spectra of galaxies, and 10,000 quasars over 10000 $\rm deg^{2}$ of the sky. The SDSS DR7 provides such a large amount of samples to study the physical properties of the local universe. The optical emission line data are from SDSS DR7 \citep{2009ApJS..182..543A} because this release consists of nearly 1 million spectra, which spans a wavelength coverage from 3800 \AA \ to 9600 \AA \ with a resolution of nearly 2000. Due to the reason that the diameter of each fiber is $3''$, we only analyze the central region of galaxies. The redshift range of the galaxies we used in this paper is from 0.02 to 0.25, thus the size of the central region is from 1.36 kpc to 14.86 kpc. Most have data within 5 kpc.

	\subsection{Star Forming Galaxies and Passive Galaxies}
	We use strong optical emission lines to calculate the metallicity, and remove the contributions from AGN. Therefore, we focus on the star forming galaxies. The BPT diagram \citep{1981PASP...93....5B} is exploited to select the star-forming galaxies, which requires that galaxies fall below the $\rm [NII]\lambda 6584/H\alpha$ versus $\rm [OIII]\lambda 5007/H\beta$ diagram \citep{2001ApJS..132...37K,2003MNRAS.346.1055K}. The optical emission lines, the stellar mass, the specific star formation rate (sSFR), and star formation rate (SFR)  \citep{2004MNRAS.351.1151B} for SDSS DR7 galaxies are from the MPA-JHU\footnote{\url{https://wwwmpa.mpa-garching.mpg.de/SDSS/DR7/}} catalog. After removing galaxies with unreliable emission lines ($\rm SNR \leq 5$), we found 68355 star-forming galaxies in the SDSS DR7 catalog. We use the definition of passive galaxies ($\rm sSFR \leq 10^{-11} \ year^{-1}$) from \cite{2011Sci...334..948T}.

	\subsection{Pair Galaxies and Match to Controls}

	Pair galaxies are selected using the following criteria: 
	the projected separation between galaxies need to be  less than 300 kpc and larger than 7.21 kpc \citep{2008ApJ...685..235P}; the line-of-sight velocity difference $\rm |\Delta V| < 1000 \ km/s$ \citep{2019ApJ...874...18W}; at least one neighbor was observed by SDSS DR7. The minimum separation is used to remove the case that the very close galaxy pairs are mistakenly classified as a single galaxy \citep{2008ApJ...685..235P}.

	\begin{table}
	\centering   
	
	\caption{Conditions of different types of pairs}
	\begin{tabulary}{0.47\textwidth}{CCC}  
	
		\hline
		\hline
		Type of pairs&\multicolumn{2}{c}{Projected  distance}\\
			%\cline{3-5}
			 &$\rm less \ than \ 150 \ kpc$ & $\rm between \ 150 \ kpc \ and \ 300 \ kpc$ \\
		\hline
        SF-SF pairs & interaction and CGM sharing & no interaction but with CGM sharing \\ 
        \hline
        SF-Passive pairs & interaction and without CGM sharing & no interaction and without CGM sharing  \\ 
		\hline
		
	\end{tabulary}
	\label{condi_diff}
\end{table}

	This work is going to study the CGM surrounding galaxies in pairs, not only the interaction between galaxies. Therefore as long as two galaxies share some overlap of their CGMs, we can use them no matter they are interacting with each other or not. Since \cite{2011Sci...334..948T} detected CGM out to 150 kpc for a galaxy and \cite{2020arXiv200808092W} found that the radius of CGM is larger than 150 kpc, so we can select a galaxy pair with a separation smaller than 300 kpc. Here we assume that galaxies with separation smaller than 300 kpc have somewhat overlapped CGMs. We finally find 4297 SF-SF pairs and 12254 SF-Passive pairs. The SF-passive pairs only have interaction but no CGM sharing, which will be used to understand the effect of CGM sharing in the SF-SF pairs.
	
\begin{figure}
		\centering
		\begin{multicols}{2}    
			\includegraphics[width=0.5\textwidth]{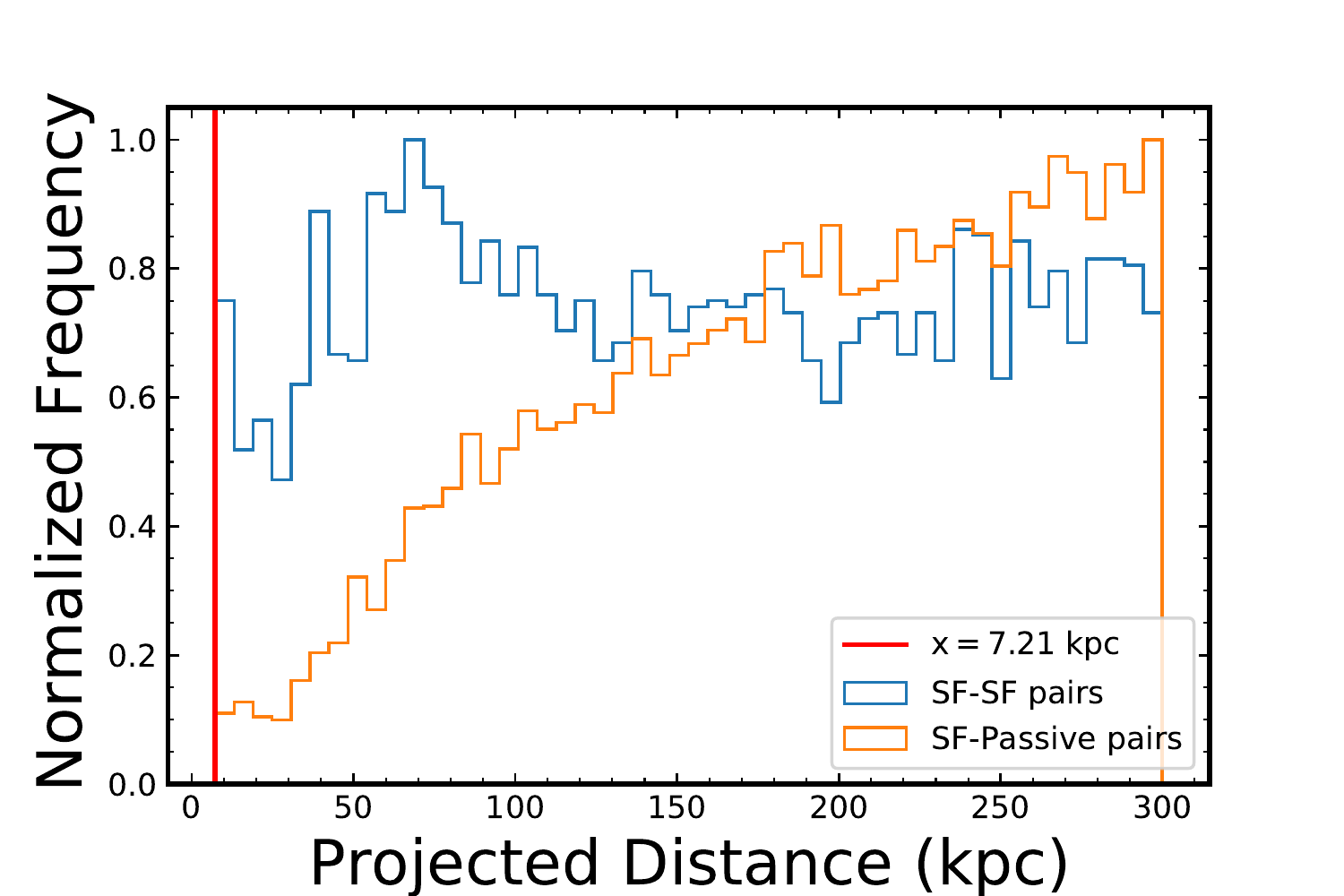}
			\includegraphics[width=0.5\textwidth]{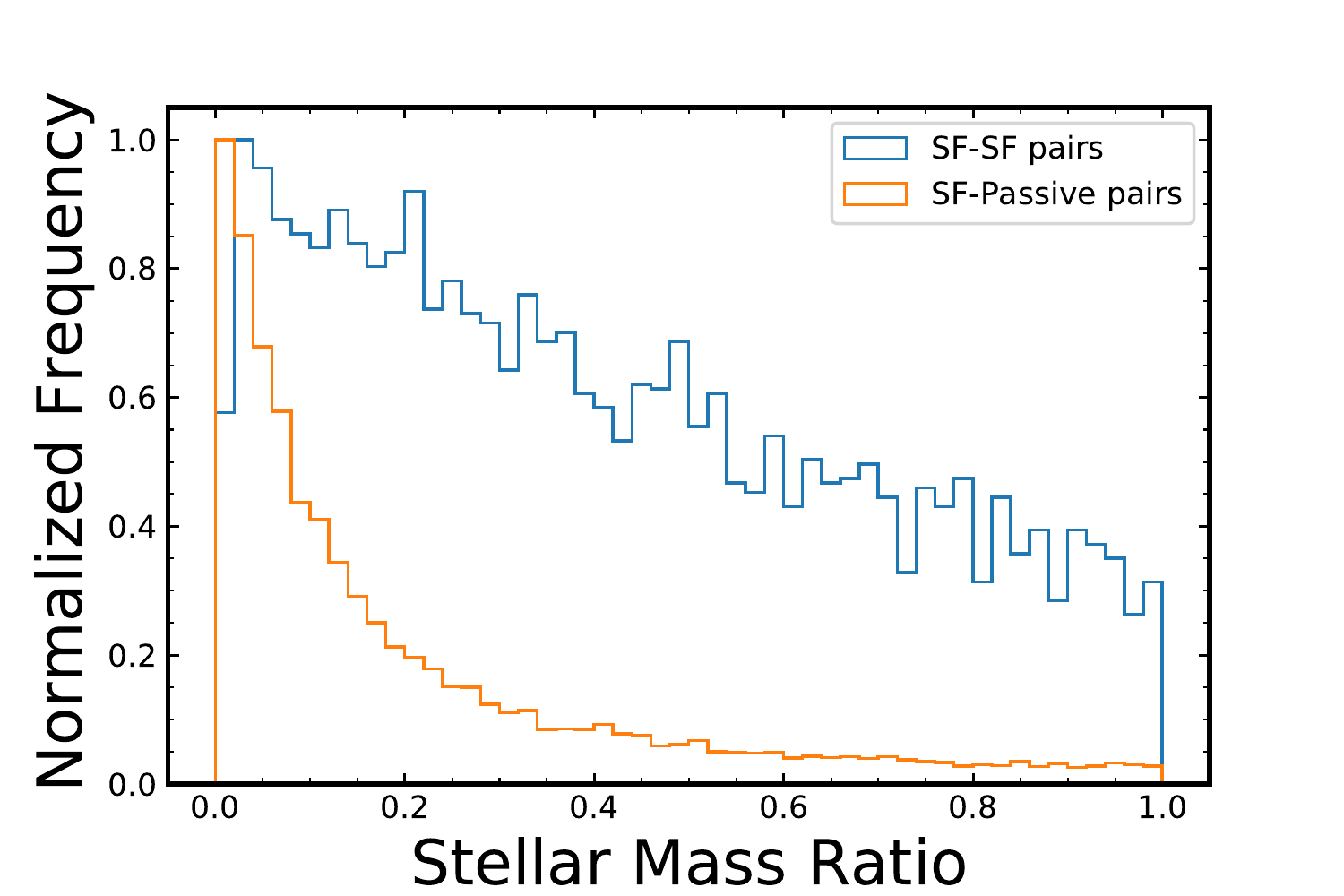}
		\end{multicols}    
		\caption{The normalized frequency distribution of the separation and the stellar mass ratio for pairs from the SDSS DR7 are shown in this figure. The red vertical line in the top panel represents the projected distance between pairs equals to 7.21 kpc. The blue, orange color show the SF-SF pairs, SF-Passive pairs, respectively.}
		\label{pair}
	\end{figure}

	The projected distance distribution and the stellar mass ratio are shown in Figure \ref{pair}. The red vertical line in the top panel is used to guide our eyes to show that we indeed remove the pairs with separation less than 7.21 $\rm kpc$. For both SF-SF pairs and SF-Passive pairs, when the projected distances of pairs are smaller than $\rm 100 \ kpc$, there are more pairs with increasing separation. However, for SF-SF pairs, the separation seems to be evenly distributed when it is larger than $\rm 100 \ kpc$. Most of the SF-SF pairs have equal stellar masses. But for SF-Passive pairs, most of them are unequal in stellar masses. 
	
 After removing pair galaxies from the whole sample of star forming galaxies, we have 24586 isolated galaxies left for the control pool. {\color{black} Following the method in \cite{2016MNRAS.461.2589P}}, we select five control galaxies from the control pool with similar stellar mass (0.1 dex) and redshift (0.01) for each star forming galaxy in pairs. 
Because of the span of these two parameters, a weighting scheme is needed to quantify the matches of pairs and control samples. Thus, two parts should be taken into account when calculating the overall weight: one for redshift weight and another for stellar mass weight. The $i-th$ overall weight of the control sample is defined as: 	
	\begin{equation}
	w_{i}= w_{z_{i}} * w_{M_{i}},
	\end{equation}
	
	where $w_{z_{i}}$ and $w_{M_{i}}$ represent the redshift weight and the stellar mass weight, respectively. The redshift weight is defined as:
	\begin{equation}
w_{z_{i}}=1-\frac{|z-z_{i}|}{z_{tol}},
	\end{equation}
	
	where $z$, $z_{i}$ and $z_{tol}=0.01$ are the redshift of pair galaxy, the redshift of $i-th$ control samples of each pair, and the redshift tolerance. The stellar mass weight is defined as:
	\begin{equation}
	 w_{M_{i}}=1-\frac{|log(M)-log(M_{i})|}{M_{tol}},
	\end{equation}
	where $M$, $M_{i}$ and $M_{tol}=0.01$ are the stellar mass of pair galaxy, the stellar mass of $i-th$ control samples of each pair, and the stellar mass tolerance.

	\subsection{Balmer Decrement and Dust Extinction Correction}
	
	Dust extinction will occur when the light comes through the dust. The scatter and absorption of dust follow this rule: the light with a shorter wavelength will be extincted more than that with a longer wavelength \citep{1990ARA&A..28...37M}. Generally, the Balmer decrement, which is the higher-order Balmer emission line flux relative to $\rm H\beta$ \citep{2005ApJ...635..260S}, is accepted to correct this effect. In this paper, we choose the ratio of $\rm H\alpha$ and $\rm H\beta$. In the Case B situation, where the temperature is $10^4 \ \rm K$, the theoretical $\rm H\alpha / \rm H\beta$ is 2.86 for the electron densities $n_{e}=10^{2} \ \rm cm^{-3}$ \citep{groves2012balmer}. Because two emission lines are close in wavelength, the difference of dust emission is so small that we can ignore it when selecting star forming galaxies. Therefore, with the Balmer decrement, the equation between the intrinsic flux and the observed flux, the reddening curve from \cite{2000ApJ...533..682C}, we obtain the intrinsic flux for other optical emission lines of star forming galaxies, such as $\rm [NII]\lambda 6584$, $\rm H\alpha$, $\rm [OIII]\lambda 5007$, $\rm H\beta$.

	\subsection{Metallicity Indicator} 
 The most reliable way to obtain metallicity is to directly measure the electron temperature of gas \citep{2008ApJ...681.1183K}. They use the ratio between $[\rm O \uppercase\expandafter{\romannumeral3}]\lambda4363$ and other low ionization emission lines such as $[\rm O \uppercase\expandafter{\romannumeral3}]\lambda5007$. However, $[\rm O \uppercase\expandafter{\romannumeral3}]\lambda4363$ usually exists in the metal-poor region of the galaxies and it is too weak to be observed \citep{garnett2004first}. Thus, other indicators are introduced to make up for these problems. For example, the classical method is based on the photoelectric model, and the method is based on the combination of the electron temperature and the photoelectric model. However, the photoelectric model is only suited for the geometrical spherical situation, which can not exist in the real universe environment \citep{2008ApJ...681.1183K}. Therefore, in this paper, we choose two metallicity indicators and cross-check the results.
	
One indicator from the photoionization model is KD02
\citep{2002ApJS..142...35K} .
It depends on the relationship between $\rm R_{23}=([O\uppercase\expandafter{\romannumeral2}]\lambda3727+[O\uppercase\expandafter{\romannumeral3}]\lambda\lambda4959, 5007)/H\beta$ and metallicity. $\rm R_{23}$ provides an estimation of the ratio of total cooling caused by oxygen. However, for this method, $\rm [N \uppercase\expandafter{\romannumeral2}]\ / [O \uppercase\expandafter{\romannumeral2}]$ are needed to break the degeneracy between $\rm R_{23}$ and metallicity. Another metallicity indicator D02 \citep{2002MNRAS.330...69D} is derived by the monotonic relationship between logarithmic $[\rm N \uppercase\expandafter{\romannumeral2}]\lambda6584$/$\rm H\alpha$ ratio and metallicity for 155 HII regions.

	\section{Results}

\begin{figure*}
		
		\centering
		\begin{multicols}{2}
			\includegraphics[width=\columnwidth]{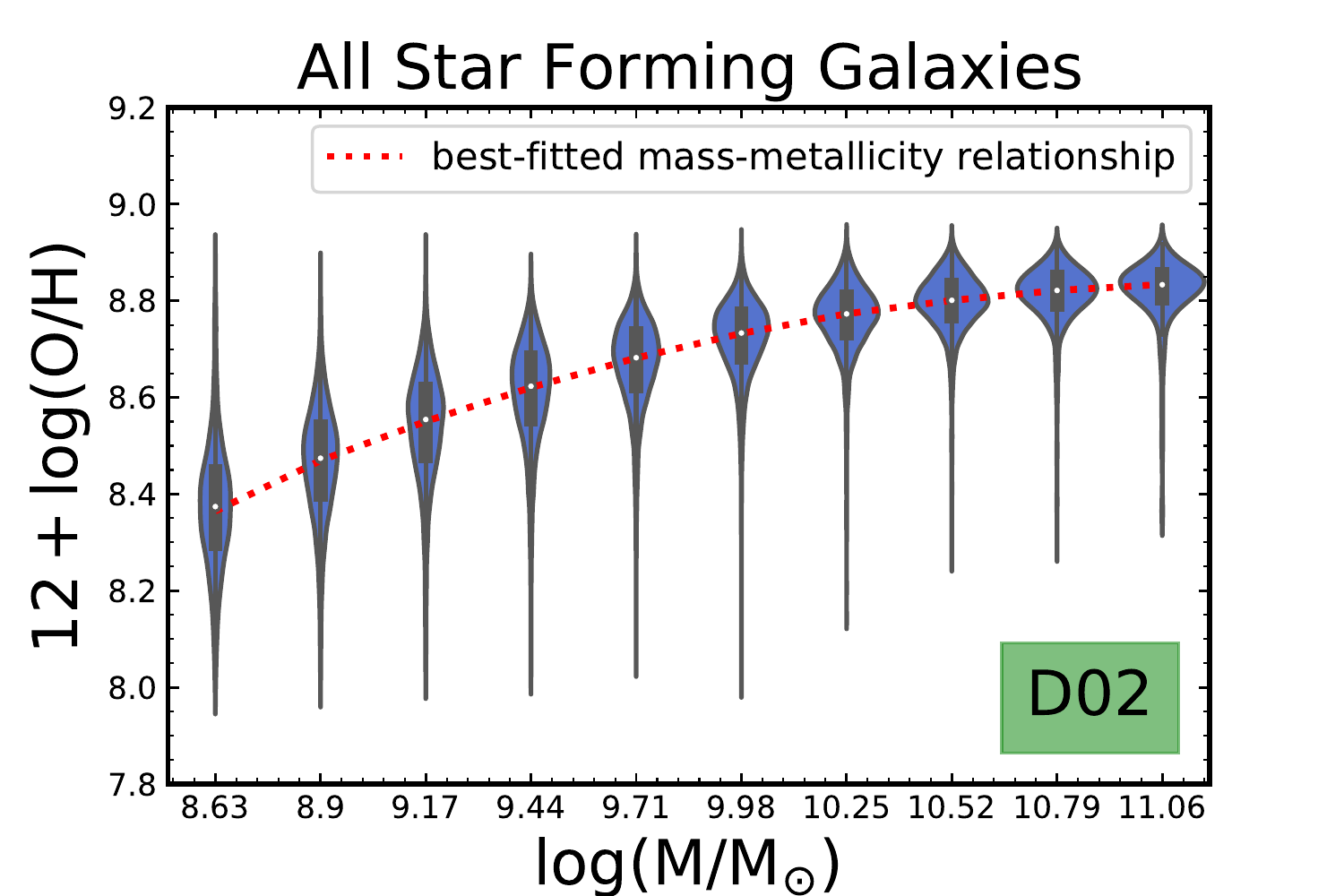}\par
			\includegraphics[width=\columnwidth]{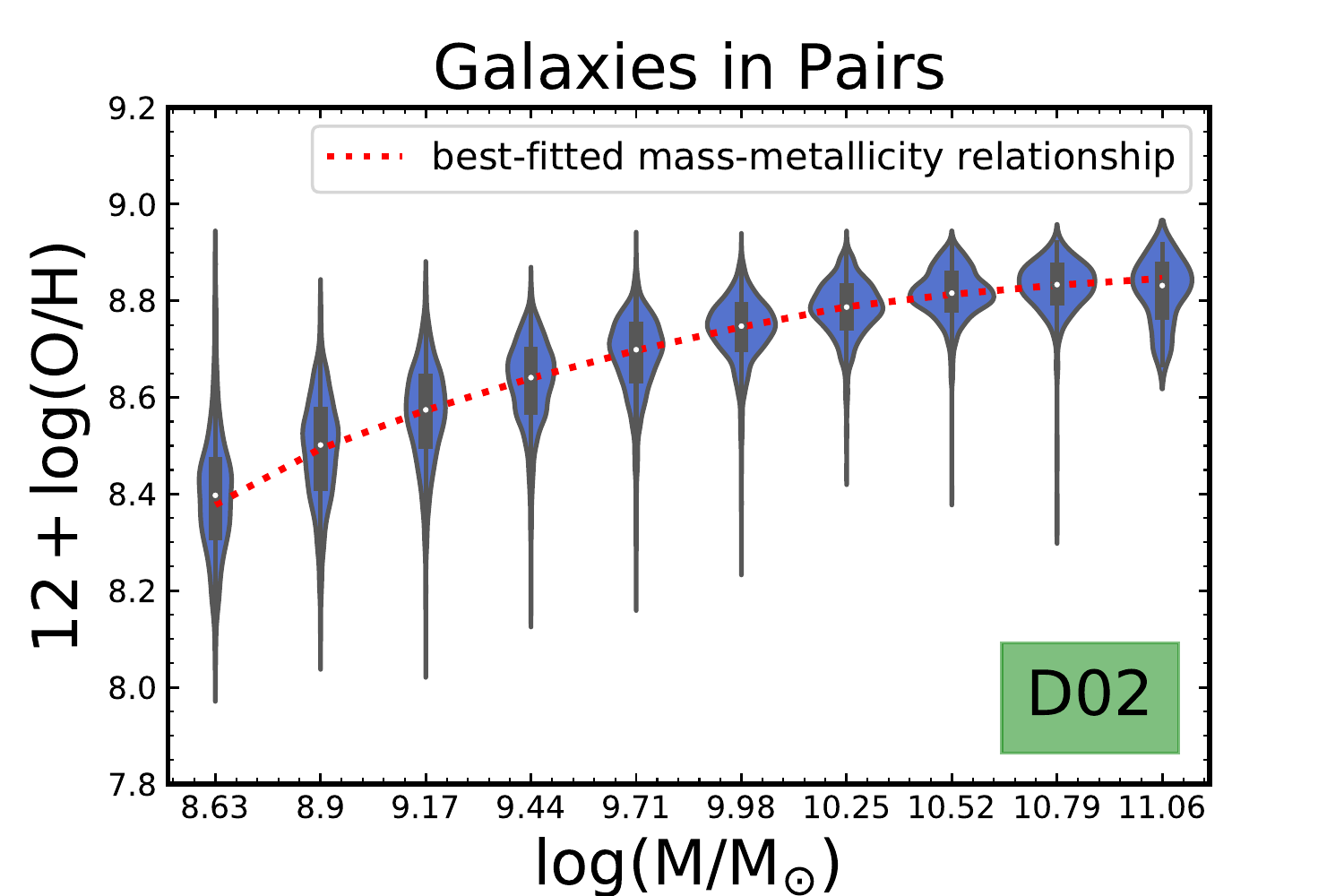}\par
		\end{multicols}
		\begin{multicols}{2}    
			\includegraphics[width=\columnwidth]{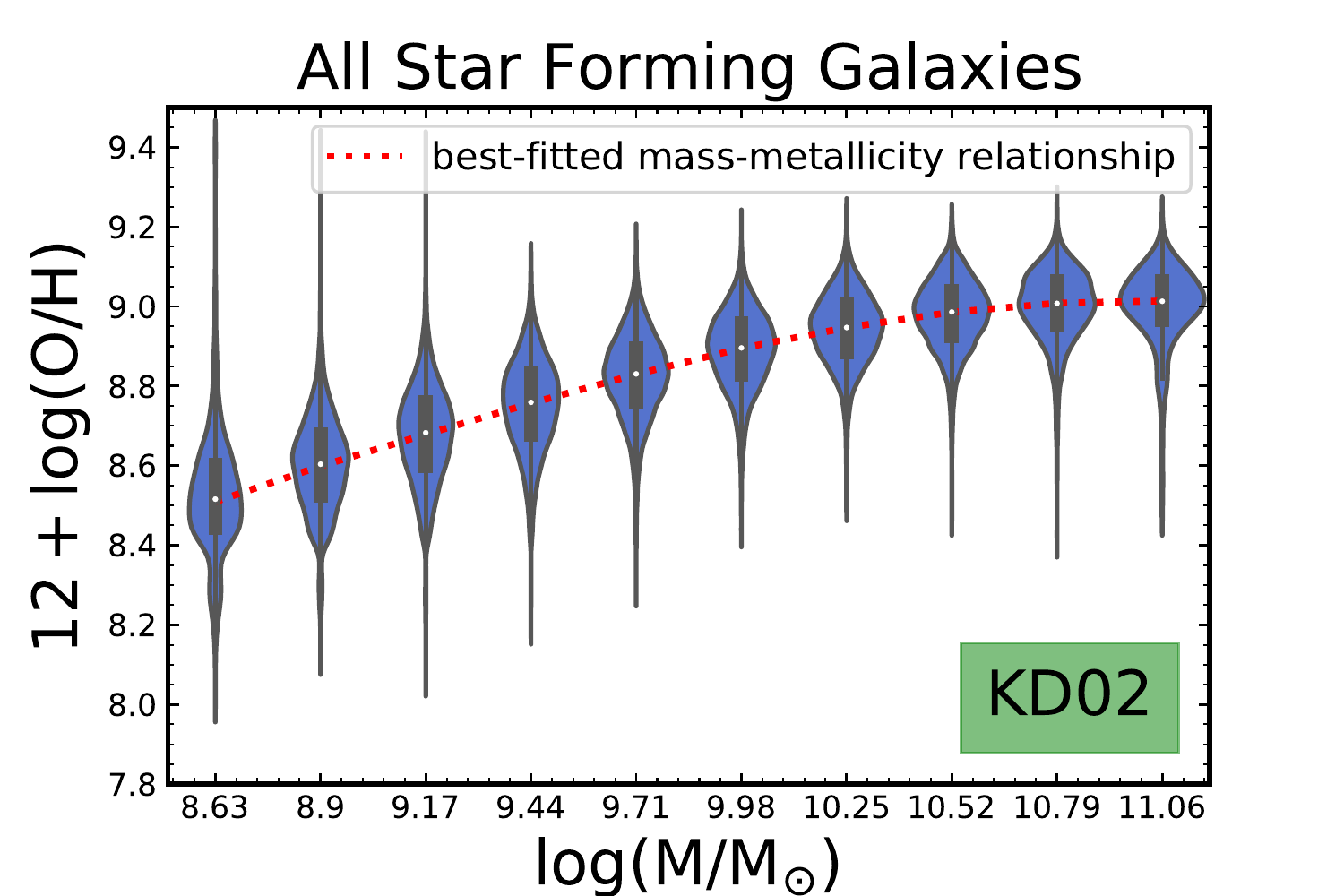}\par
			\includegraphics[width=\columnwidth]{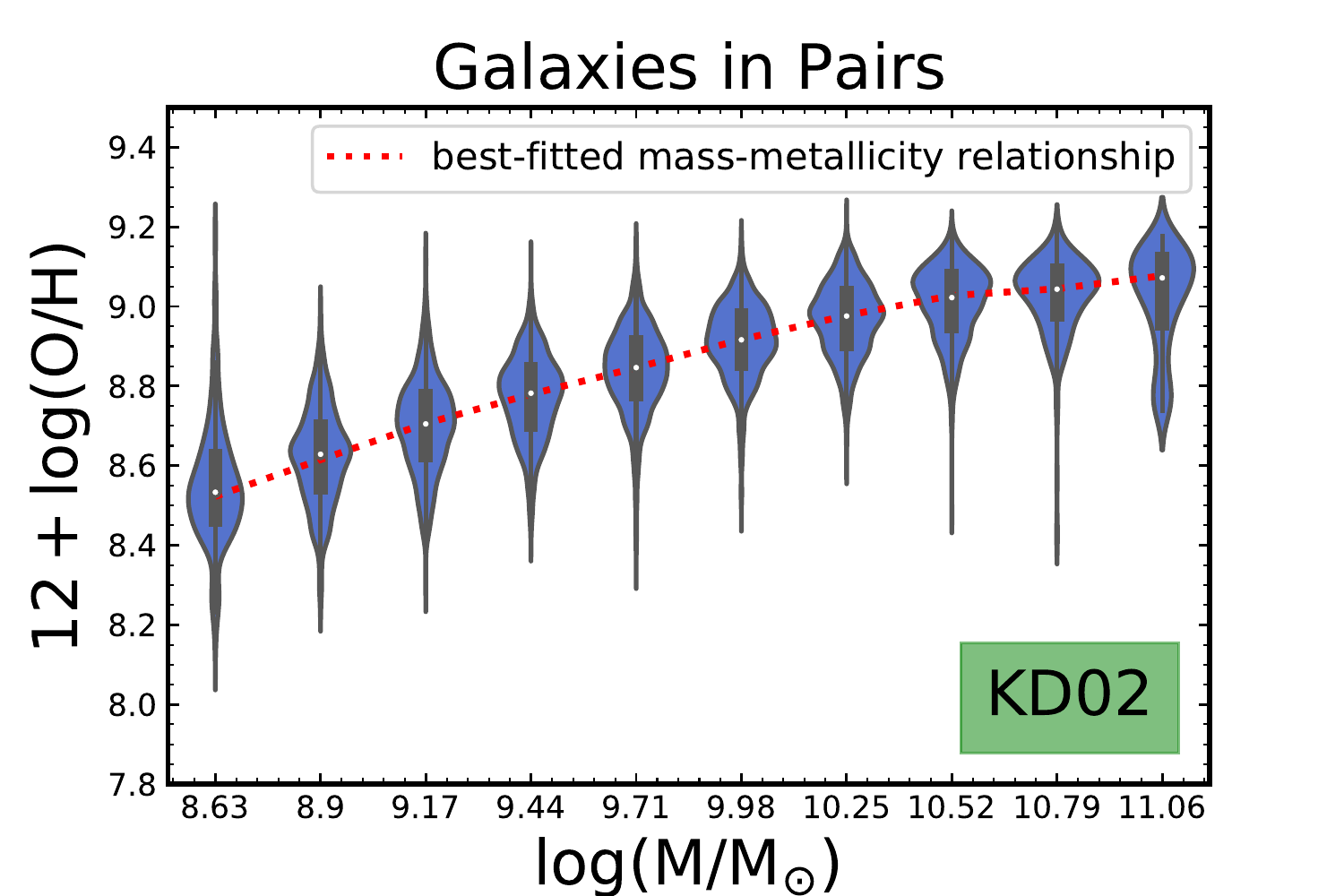}\par
		\end{multicols}    
		\caption{Violin plots illustrate the probability distribution of metallicity in each stellar mass bin in the stellar mass-metallicity relationship of star-forming galaxies from SDSS DR7. In the violin plot, the white spot, the thick black bars, the thin black bars, the thin black lines in each bin represent the median value, the interquartile range, 1.5 $\times$ interquartile range, the distribution of y-axis data, respectively. The metallicity indicator in the top, the bottom panel are from D02, KD02, respectively. The left, right panels represent the relationship for all star-forming galaxies, pairs (SF-SF pairs and SF-Passive pairs), respectively. All star-forming galaxies contains pairs and the control galaxies. The red dotted lines represent the ``best-fitted mass-metallicity relationship''.}
		\label{ratio_dr7}
	\end{figure*}

	Figure \ref{ratio_dr7} shows the stellar mass-metallicity relationship of all star-forming galaxies and pair galaxies in the violin plot. Similar to the results from \cite{2004ApJ...613..898T}, there is a tight relation between stellar mass and metallicity, showing a monotonic correlation and getting flat when stellar mass is higher than  $10^{10.5} \  M_{\odot}$. Overall, the metallicity from KD02 is larger than that of D02. This difference is introduced by the different calibration we used. The typical error of metallicity for KD02 (all star forming), KD02 (pairs), D02 (all star forming), D02 (pairs) are 0.010, 0.011, 0.009, 0.009, respectively, which are small enough to impact the final results. We divide the stellar masses into 10 bins and calculate the median value of metallicity in each bin. Then we interpolate these 10 points to obtain the best-fitted mass-metallicity relationship by using the ``scipy.interpolate.interp1d"\footnote{\url{https://docs.scipy.org/doc/scipy/reference/generated/scipy.interpolate.interp1d.html}} function from python. This relationship is shown in Figure \ref{ratio_dr7}.

	We aim to find out whether the sharing CGM of pairs will reduce the metallicity difference between them or not. Thus, we here calculate the metallicity difference between two members in a galaxy pair and compare them to two-fake pair galaxies from control samples that have the same stellar masses and redshift as the observed one but not physically associated. The metallicity difference between two members in pairs (SF-SF pairs) is simply defined as: 
	\begin{equation}
	\label{metallicity difference}
	    \rm \Delta log(O/H)_{diff}=Z_{pri}-Z_{sec},
	\end{equation}
	where $\rm Z_{pri}$ is the metallicity measured from strong emission line of galaxies in pairs (or ``fake'' pairs) with higher stellar mass and $\rm Z_{sec}$ the metallicity from the lower one. The sharing CGM likely decreases the metallicity difference between two members
in pairs because gas in two galaxies are contaminated by a common bulk of CGM. If $\rm |\Delta log(O/H)_{diff, fake}|>|\Delta log(O/H)_{diff, real}|$, we consider the share of the CGM between pairs decreases the metallicity difference between two members in pairs. 
	
\begin{figure*}

		\centering
		\includegraphics[width=\columnwidth]{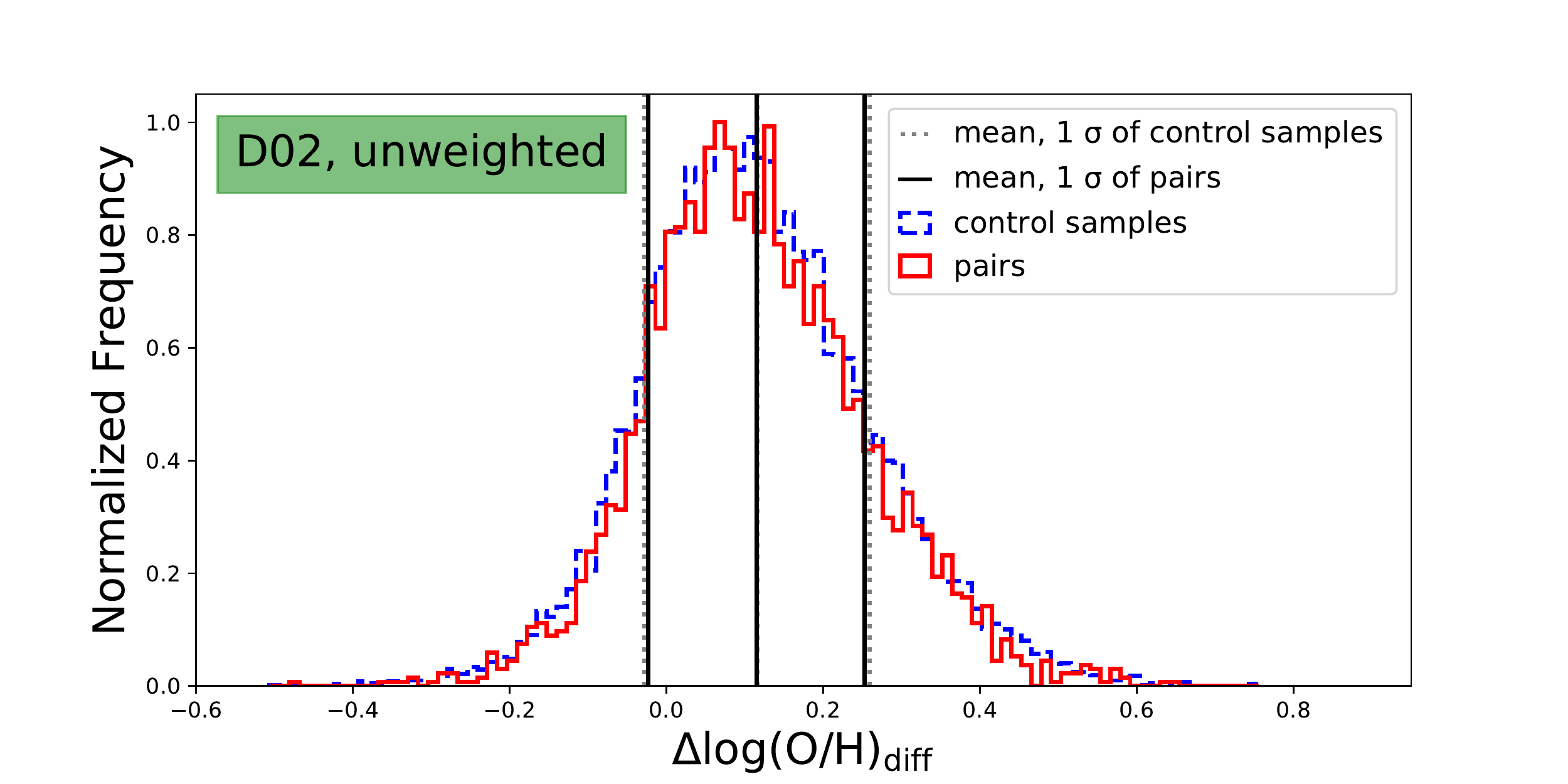}	
		\includegraphics[width=\columnwidth]{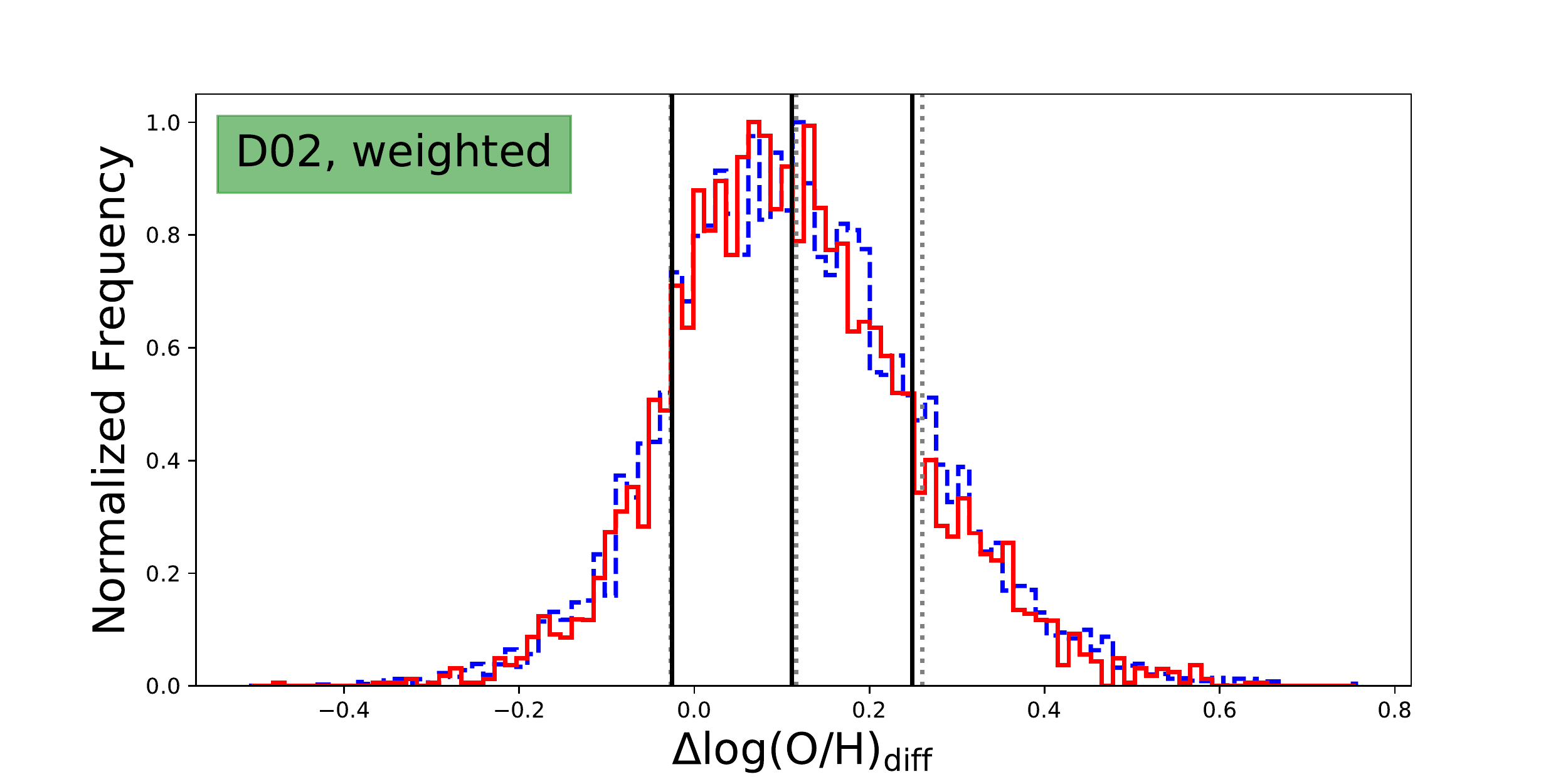}\par
		\includegraphics[width=\columnwidth]{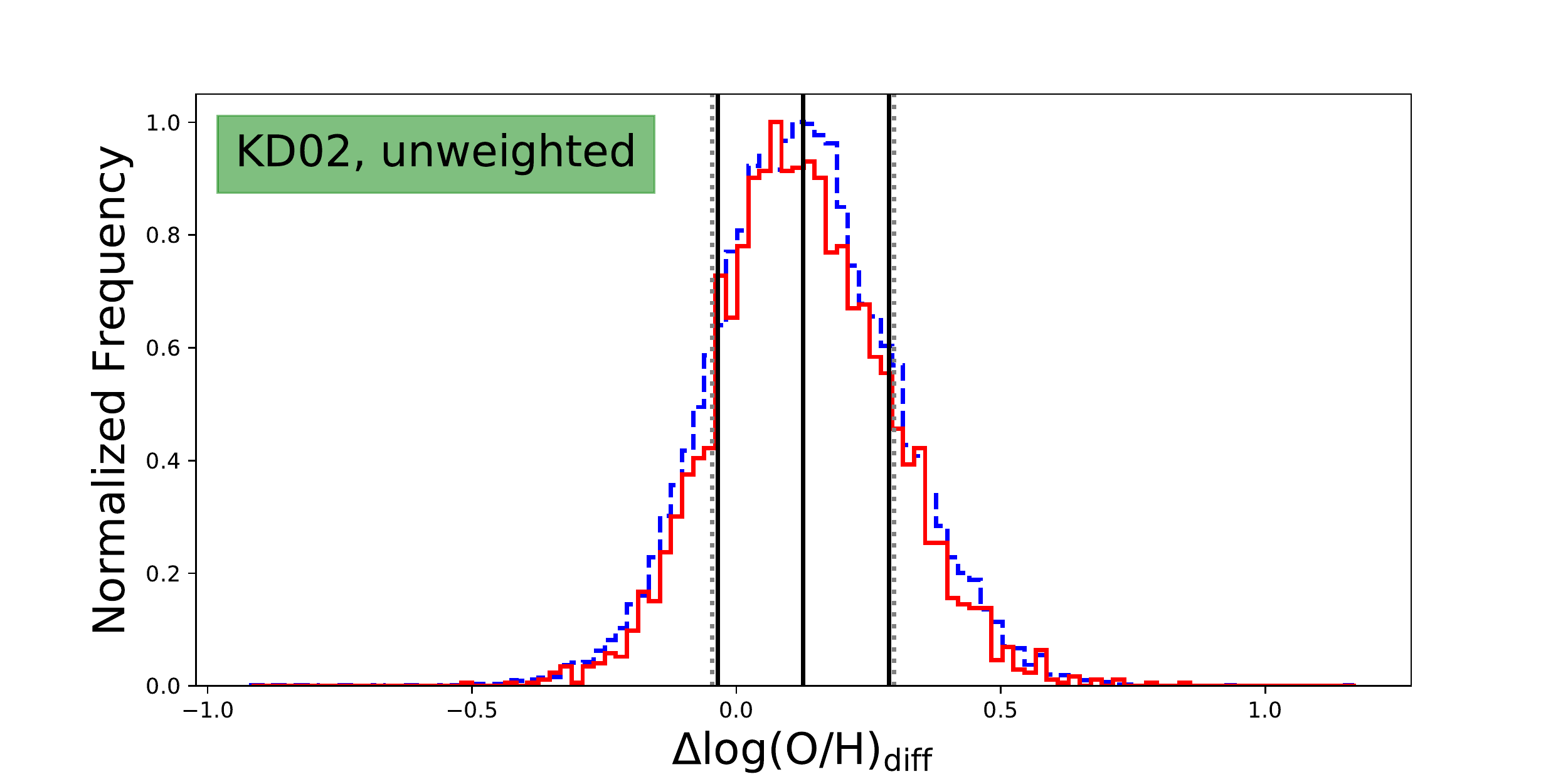}
		\includegraphics[width=\columnwidth]{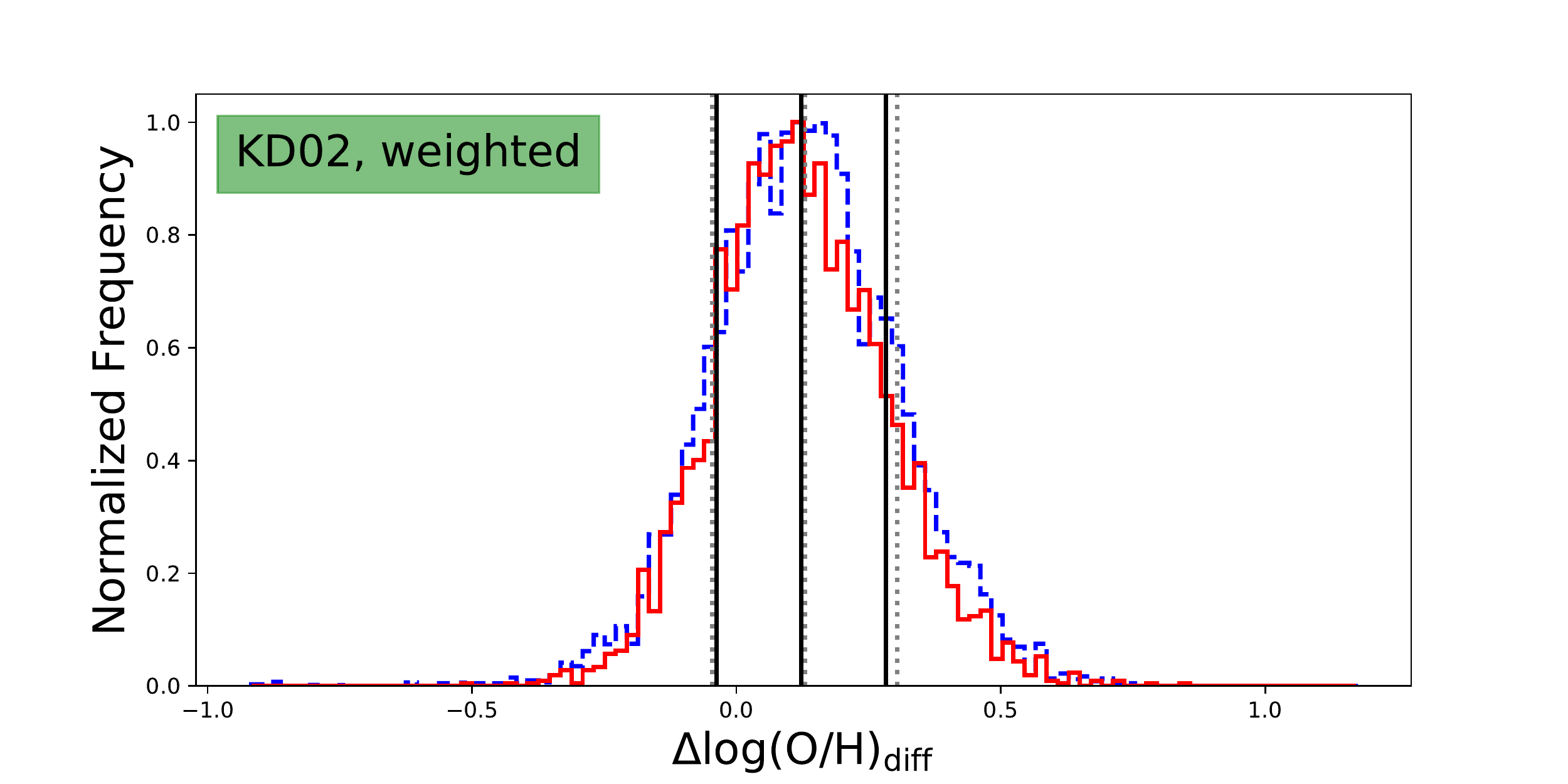} \par
			\includegraphics[width=\columnwidth]{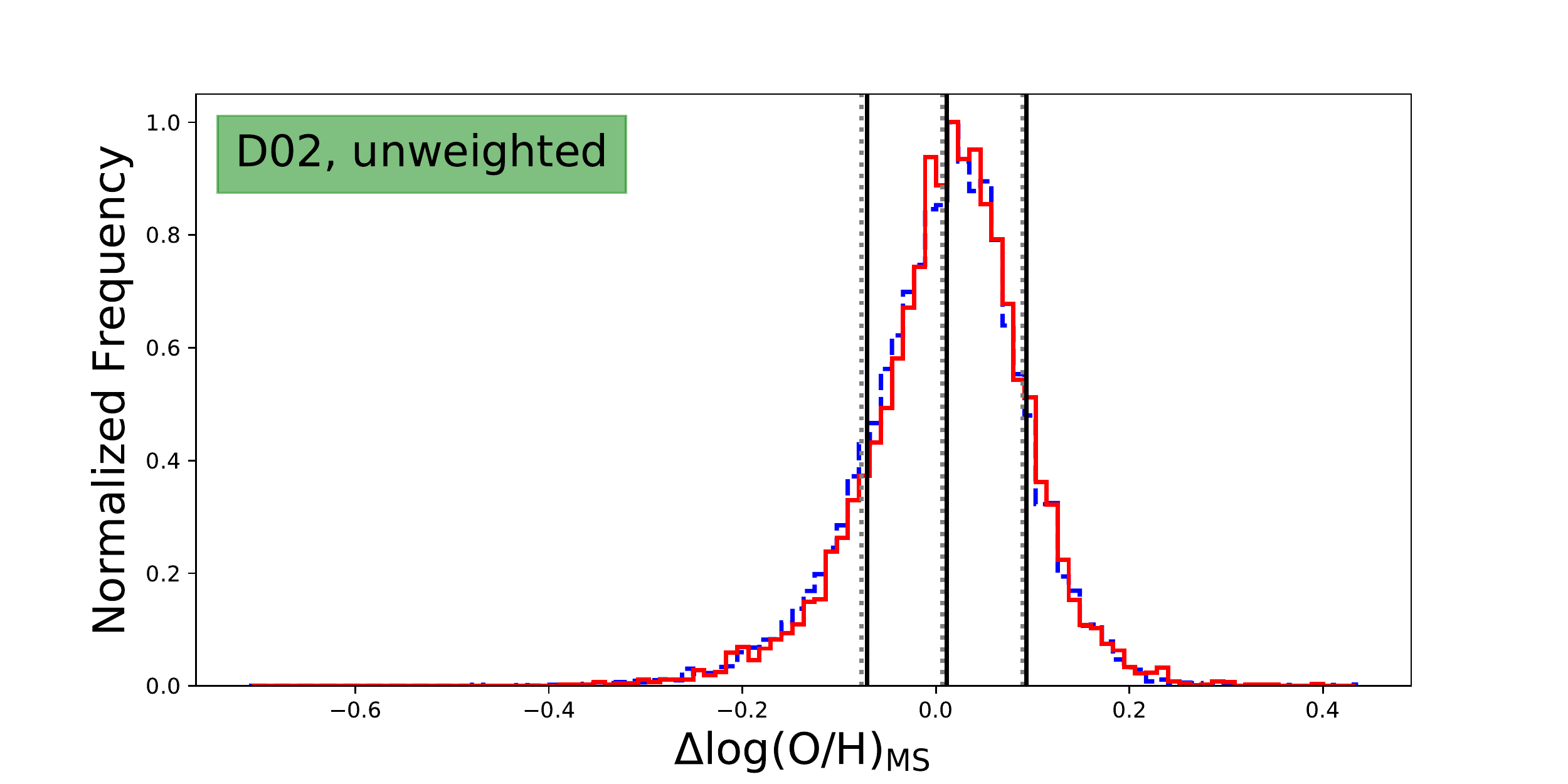}
			\includegraphics[width=\columnwidth]{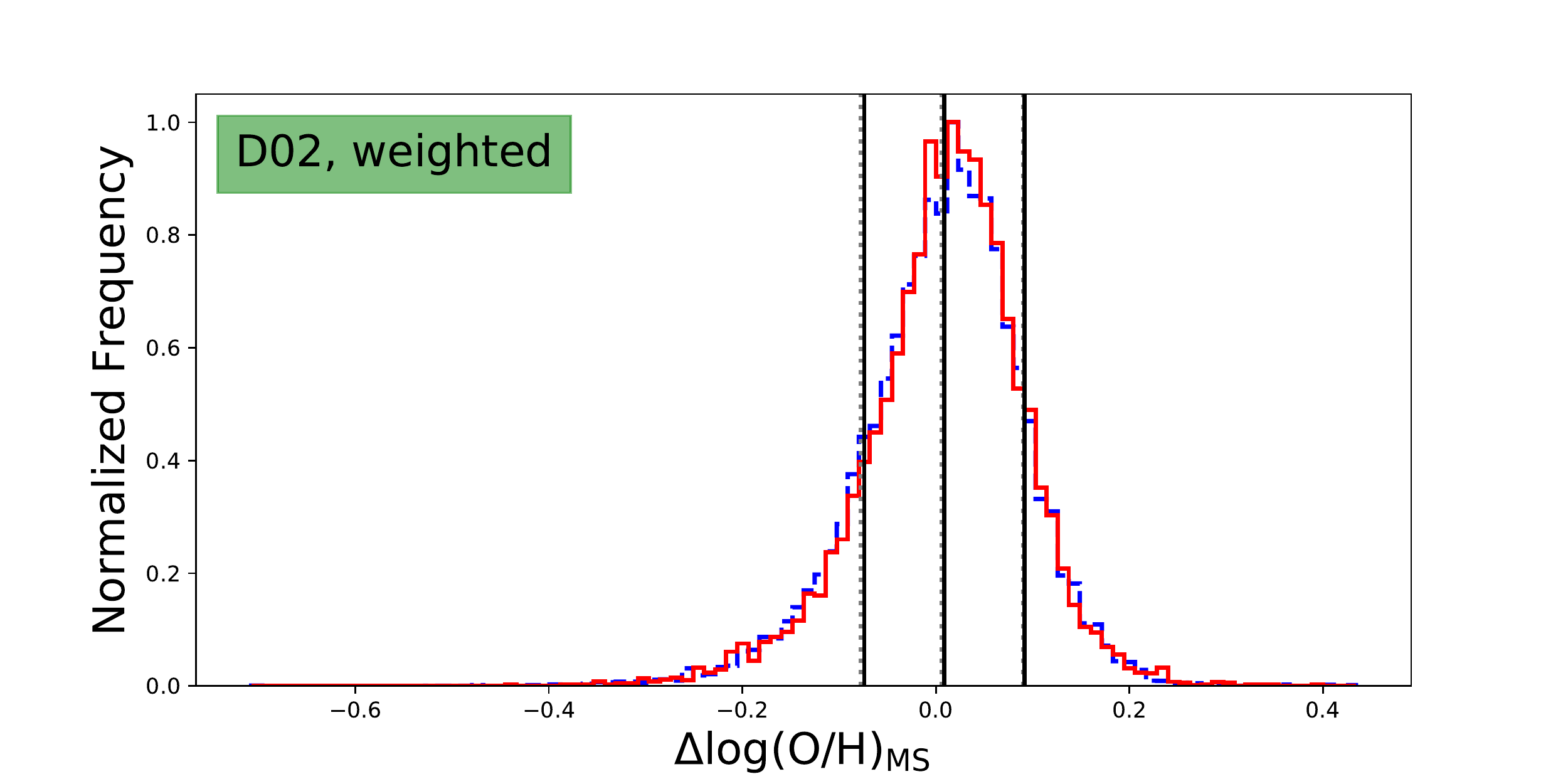}\par
			
			\includegraphics[width=\columnwidth]{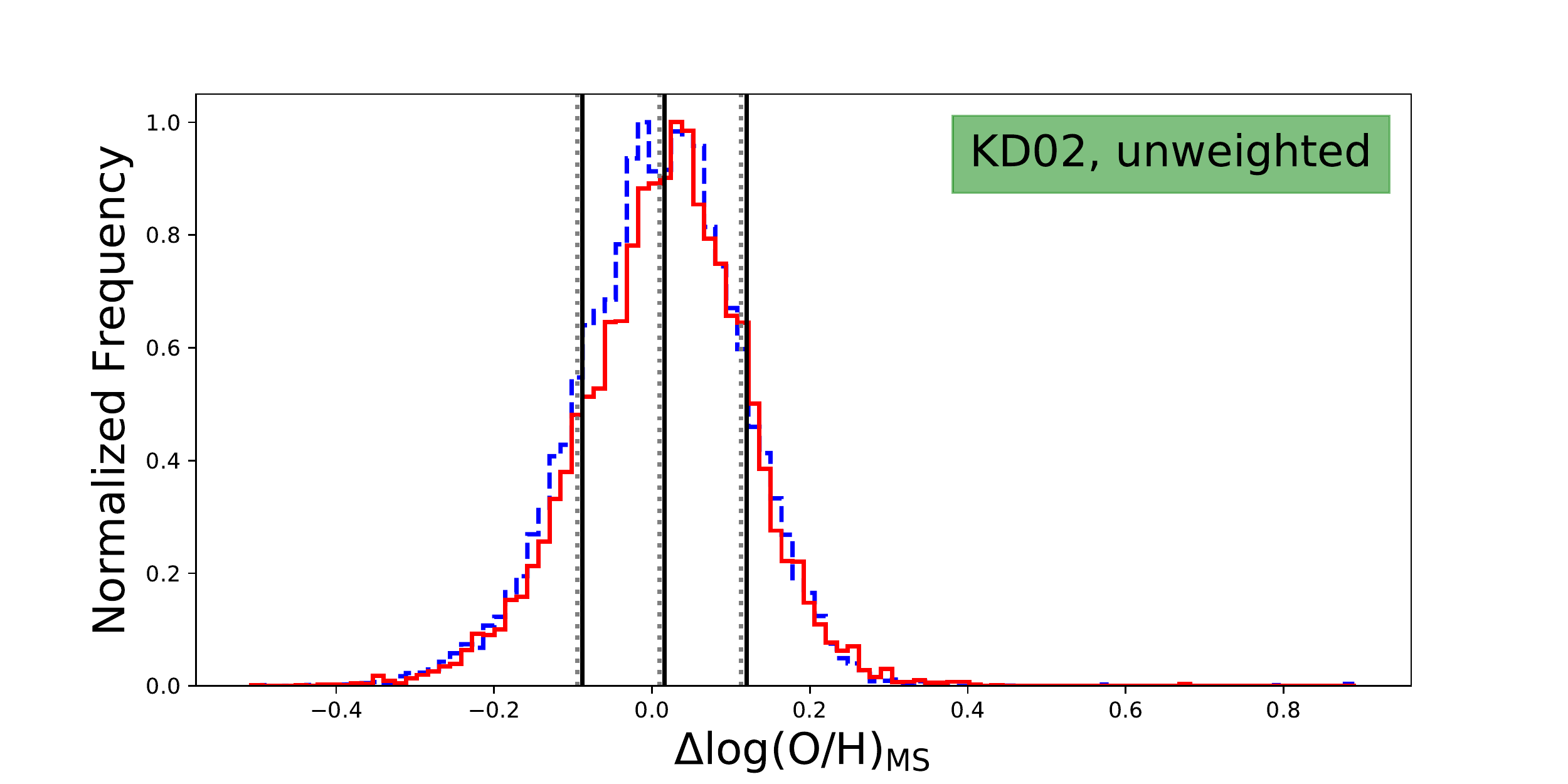}
			\includegraphics[width=\columnwidth]{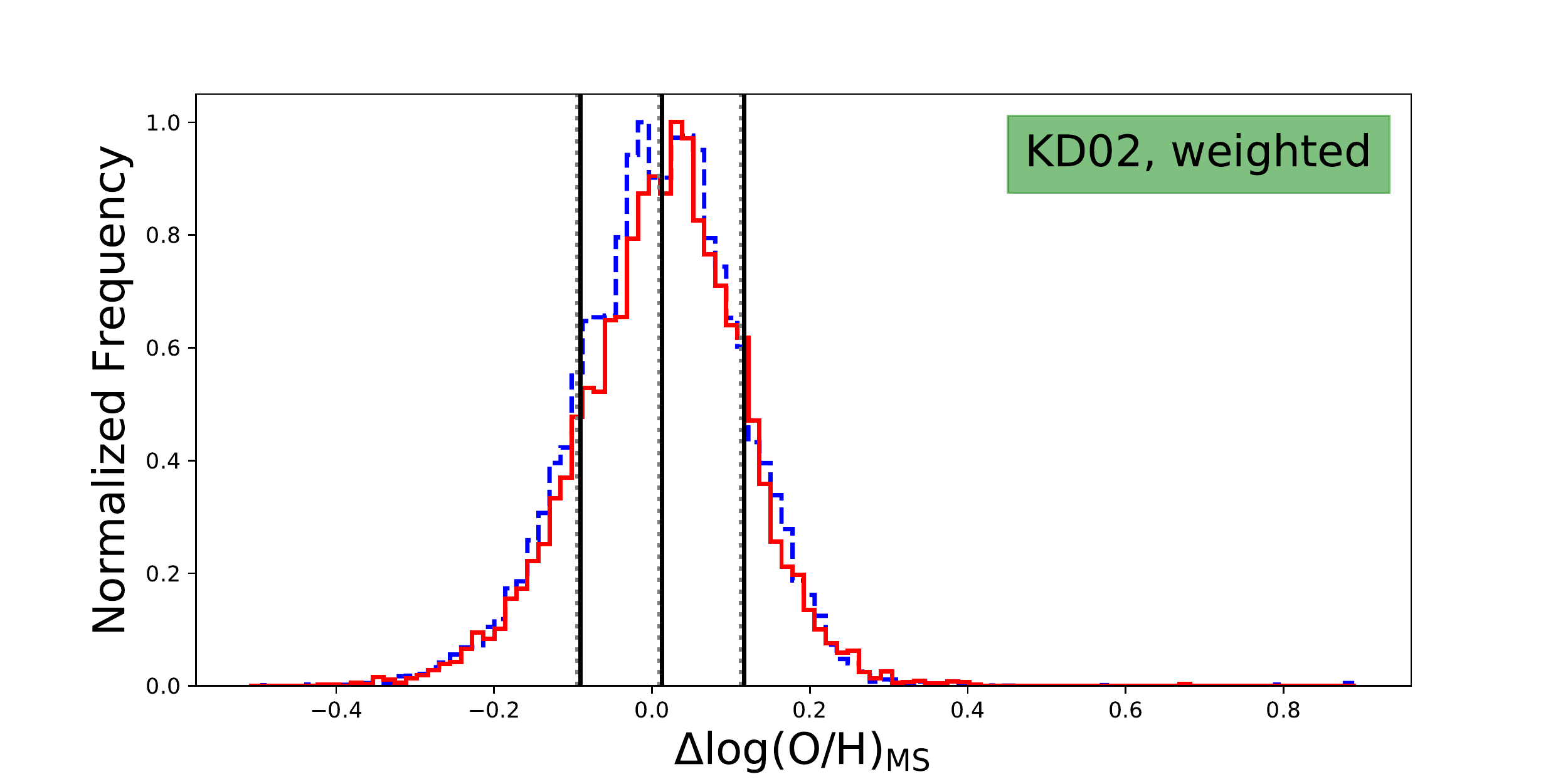}\par

			\caption{The distribution of $\rm \Delta log(O/H)_{diff}$ and $\rm \Delta log(O/H)_{MS}$ with different metallicity indicators are shown in this figure. The left panels represent the unweighted data, and the right panels represent the weight corrected data. The red solid lines are pair galaxies, the blue dashed are control galaxies. The vertical black solid line, gray dashed line illustrates the mean and $1\sigma$ of control samples, pairs, respectively. The metallicity indicator for the first and third rows are from D02, the rest of rows are from KD02.}
		\label{ratio}
	\end{figure*}

	As shown in the top 2 left panels of Figure \ref{ratio}, for both metallicity indicators, there is no significant difference in the distribution between pairs and control galaxies. Table \ref{metal_diff} provides additional statistics of those parameters. If the CGMs affect the distribution of metals between pairs, compared to the isolated fake pairs from control samples, we expect a smaller distribution difference between pairs, such as the mean and $1 \sigma$ of $\rm \Delta log(O/H)_{diff}$. However, compared to the span of the \rm $\rm \Delta log(O/H)_{diff}$ ($\approx 1.4 \rm \ or \ 2.0)$), the difference of mean 
value ($\approx 0.1150$ for D02, $\approx 0.1267$ for KD02) and $1\sigma$ ($\approx 0.1380$ for D02, $\approx 0.1622$ for KD02) between pairs and control samples are too small to be important. Thus, no significant difference is found in the $\rm \Delta log(O/H)_{diff}$ between pairs and control samples, which means that the share of CGM will not reduce the metallicity difference between pairs.

	\begin{table}
	% without passive
		\centering    
		\caption{unweighted and weighted $\rm \Delta log(O/H)_{diff}$}
		\begin{tabulary}{0.47\textwidth}{CCCCC}
			\hline
			\hline
			Metallicity Indicator&Type&Weight&\multicolumn{2}{c}{$\rm \Delta log(O/H)_{diff}$}\\
			%\cline{3-5}
			&&   &Mean & $1\sigma$ \\
			\hline
			D02 &pairs& NO &$0.1150$& $0.1380$\\
			%\cline{3-5}
			&& YES & 
			$0.1116$& $0.1373$\\
			                       
			\cline{2-5}
			&control samples&NO & $0.1157$&$0.1437$ \\
			%\cline{3-5}
			&& YES & $0.1166$& $0.1436$\\
			
			\hline
			KD02 &pairs& NO & $0.1267$& $0.1622$\\
			%\cline{3-5}
				 &     & YES &$0.1229$ & $0.1604$\\
			\cline{2-5}
		       	&control samples& NO &$0.1265$&$0.1722$\\
			%\cline{3-5}
			&& YES &$ 0.1295$ & $0.1750$\\
			\hline
		\end{tabulary}
		\label{metal_diff}
	\end{table}

		Then we analyze the metallicity offset from the best-fitted stellar mass-metallicity relationship of galaxies in pairs ($\rm \Delta log(O/H)_{MS}$), which is:
		\begin{equation}
		\label{metallicity offset}
		    \rm \Delta log(O/H)_{MS}=Z-f(M_{\star}),
		\end{equation}
		where $\rm Z$ is the metallicity of each galaxy and $\rm f(M_{\star})$ is the metallicity of a galaxy with the same stellar mass but on the main sequence. The distribution is shown in the two bottom left panels in Figure \ref{ratio}. Table \ref{offset} also displays the statistics of unweighted metallicity difference. For D02, the mean $\rm \Delta log(O/H)_{MS}$ of pairs (0.0113) is larger than that of control samples (0.0066). Overall, the difference between pairs and control samples is still too small to be significant. In this way, we conclude that there is no significant difference in $\rm \Delta log(O/H)_{MS}$ between pairs and control galaxies.

	\begin{table}
    
		\centering    
		\caption{unweighted and weighted $\rm \Delta log(O/H)_{MS}$}
		\begin{tabulary}{0.47\textwidth}{CCCCC}    
			\hline
			\hline
			Metallicity Indicator&Type&Weight&\multicolumn{2}{c}{$\rm \Delta log(O/H)_{MS}$}\\
			%\cline{3-5}
			&&   &Mean & $1\sigma$  \\
			\hline
			D02 &pairs& NO&$0.0113$ &$ 0.0825$\\
			%\cline{3-5}
			&& YES & $0.0087$ & $0.0827$\\
			
			\cline{2-5}
			&control samples& NO &  $0.0066$&$0.0834$ \\
			%\cline{3-5}
			&& YES &$0.0063$ & $0.0837$\\
			\hline
			KD02 &pairs&NO&$0.0157$&$0.1041$ \\
			%\cline{3-5}
			&& YES &$0.0130$&$0.1036$\\
			\cline{2-5}
			&control samples&NO&$0.0091$&$0.1039$\\
			%\cline{3-5}
			&& YES &$0.0092$ & $0.1039$ \\
			\hline
		\end{tabulary}
		\label{offset}
	\end{table}

\begin{figure*}
		
		\centering
		\begin{multicols}{3}
		    \includegraphics[width=0.333\textwidth]{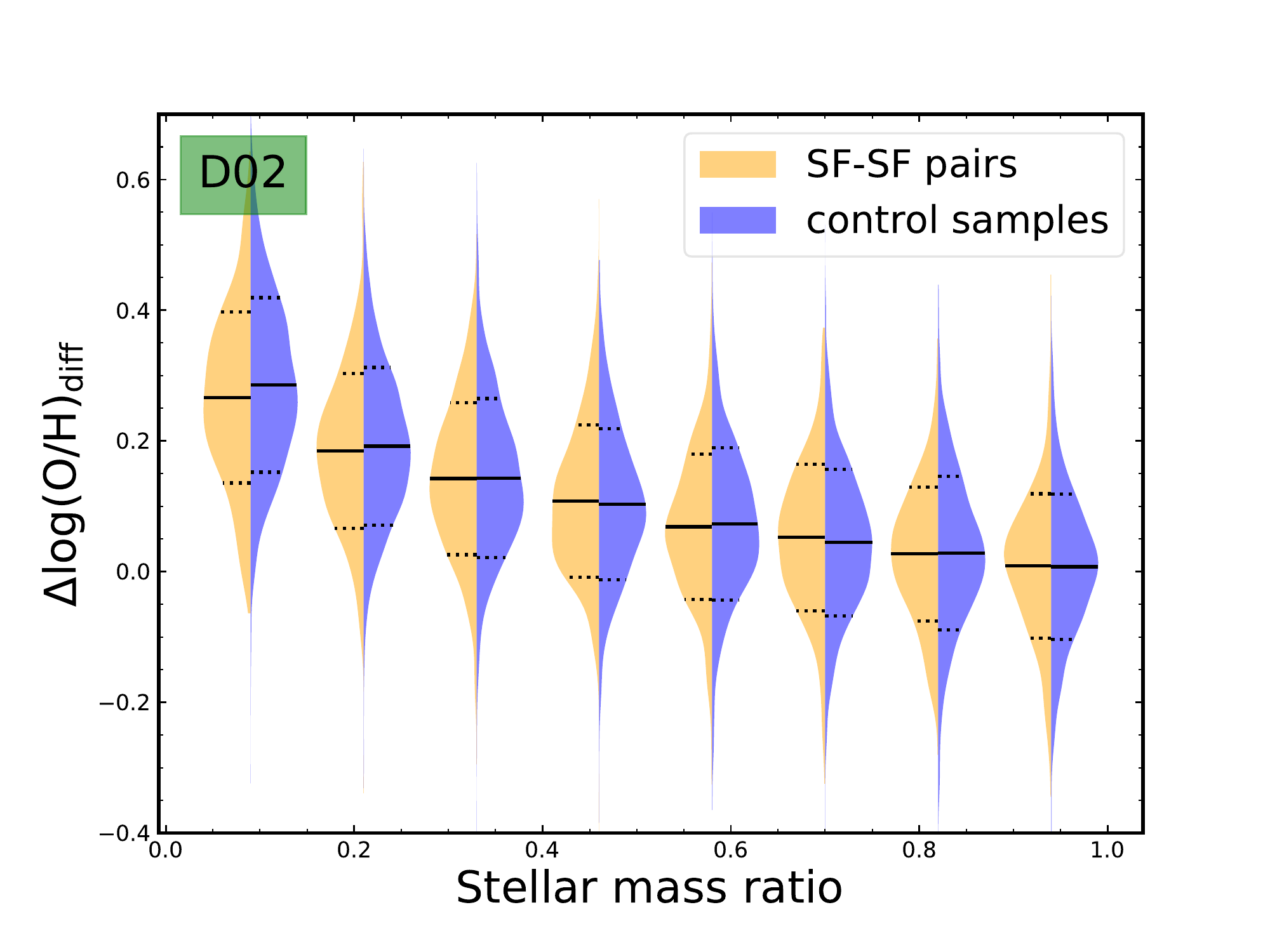}\par
			\includegraphics[width=0.333\textwidth]{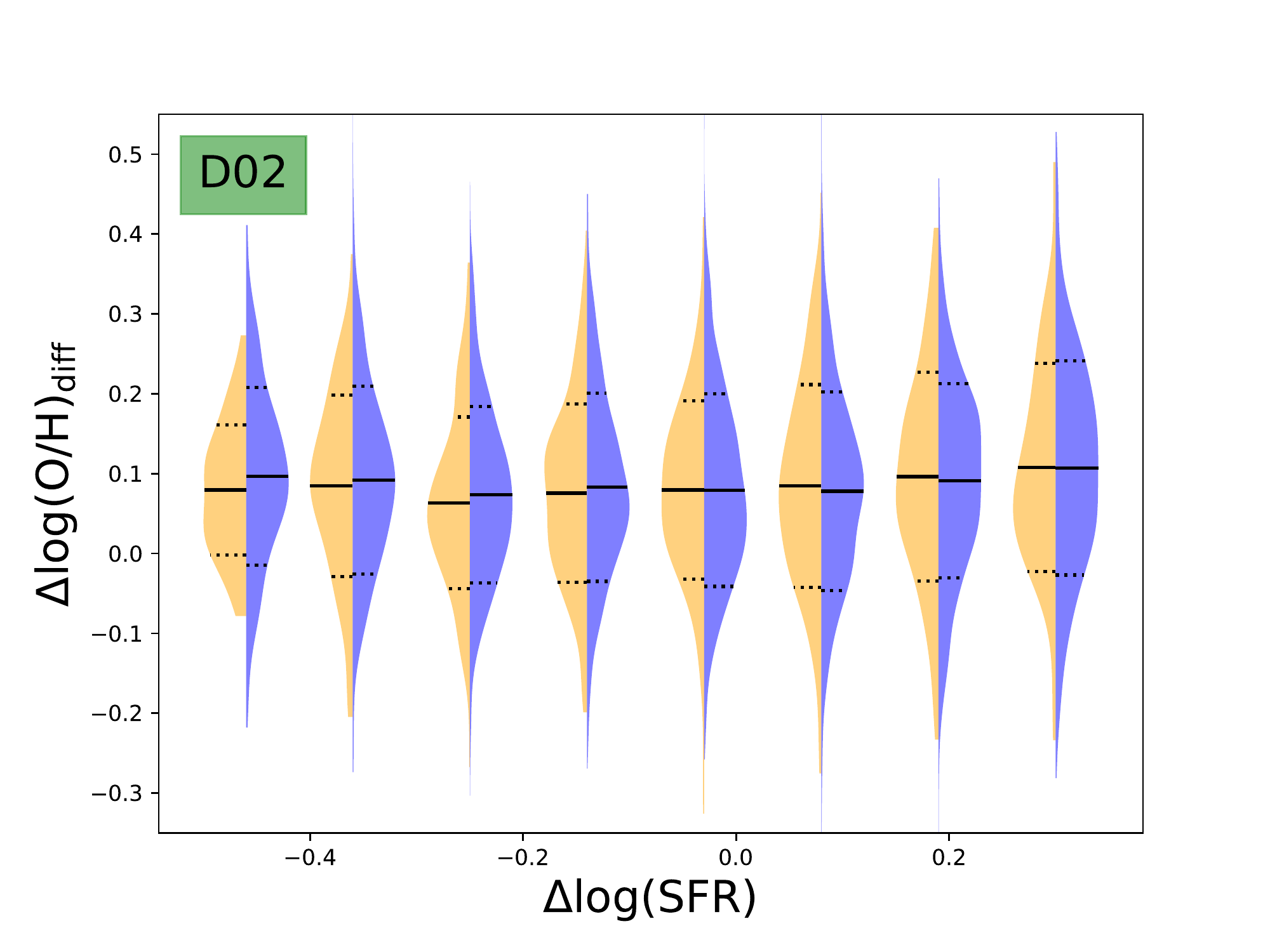}\par
			\includegraphics[width=0.333\textwidth]{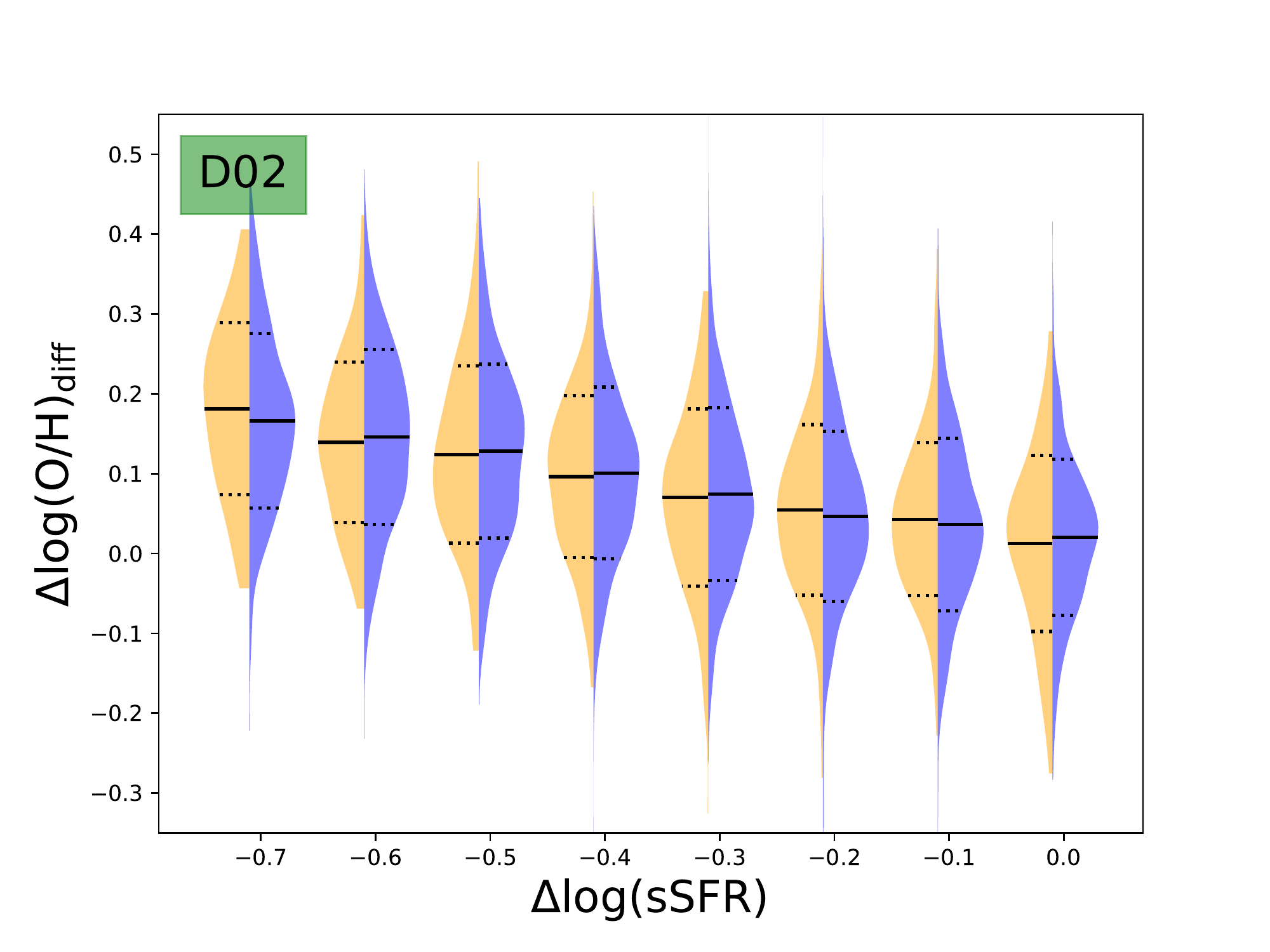}\par
			
		\end{multicols}
		\begin{multicols}{3}  
	    	\includegraphics[width=0.333\textwidth]{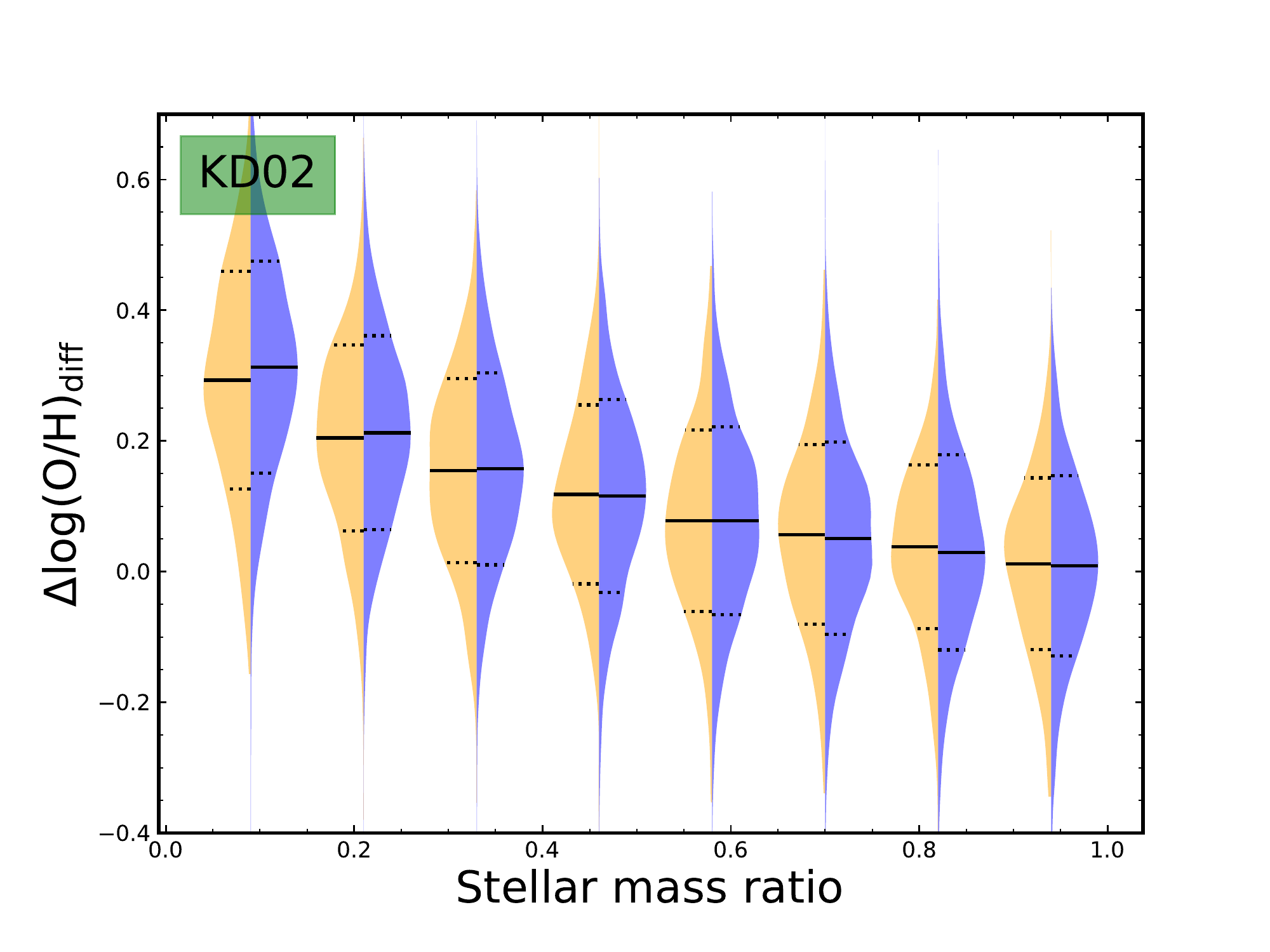}\par
			\includegraphics[width=0.333\textwidth]{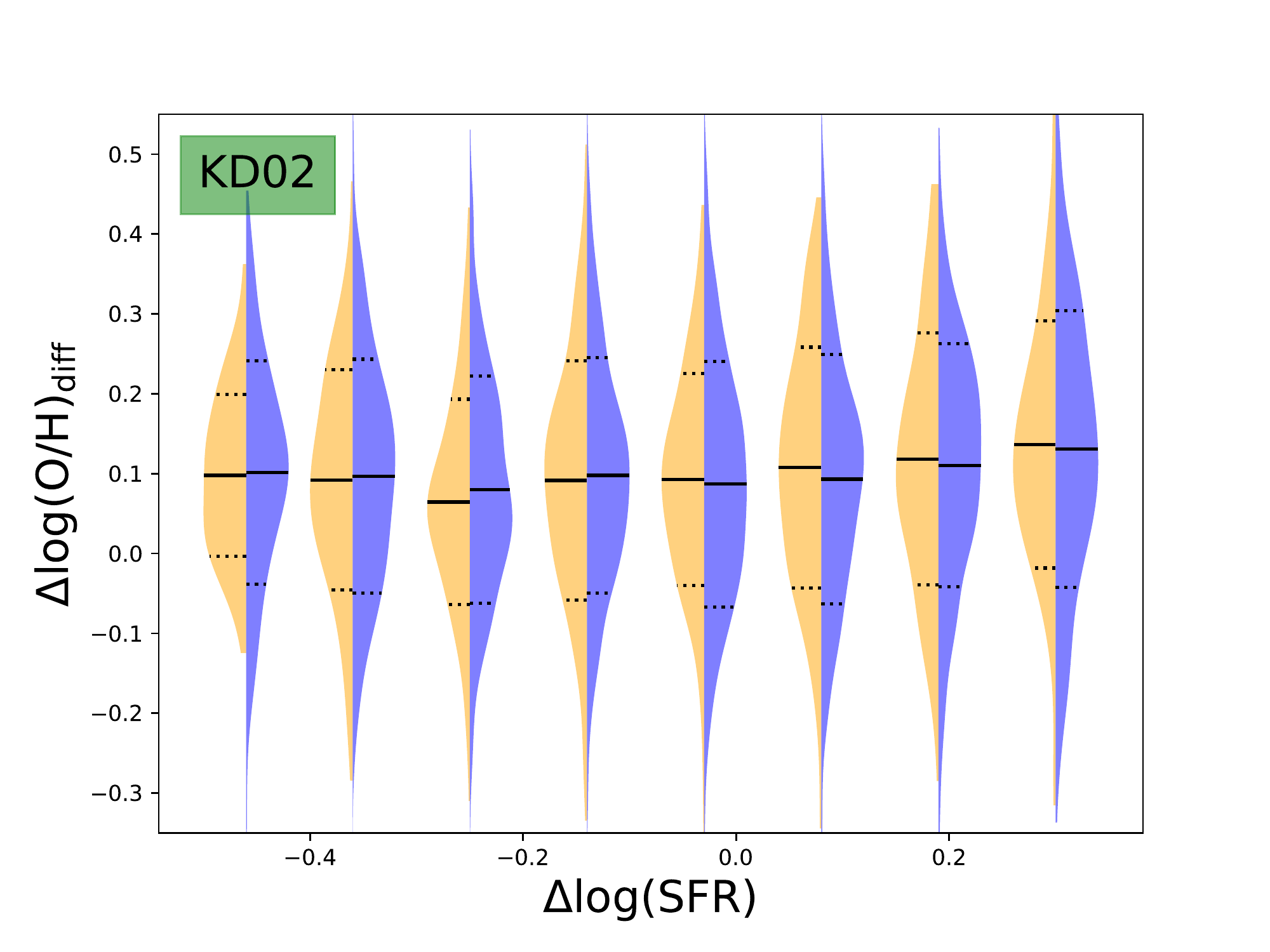}\par
			\includegraphics[width=0.333\textwidth]{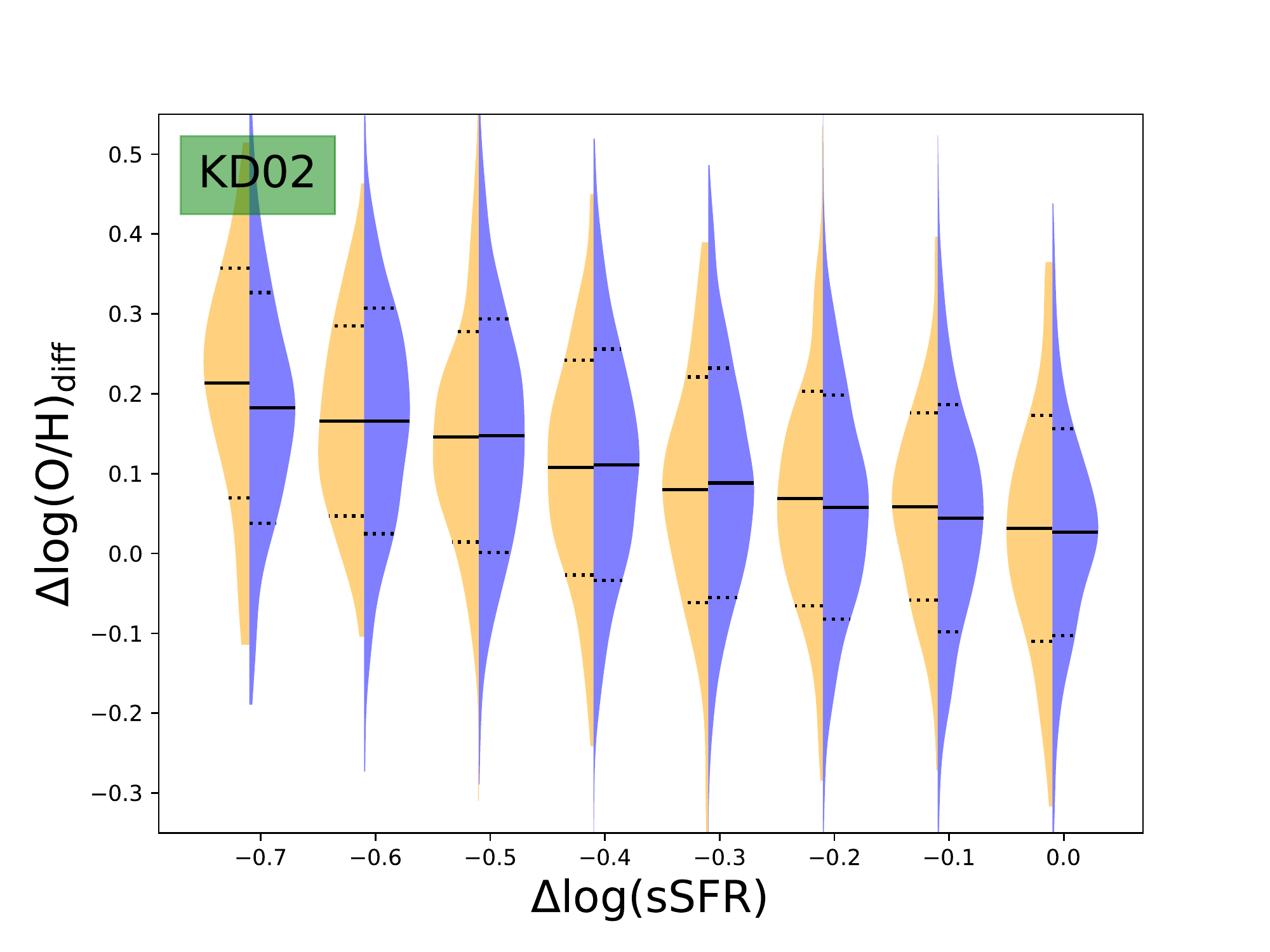}\par
			
		\end{multicols}    
		\caption{Violin plots show the probability distribution of $\rm \Delta log(O/H)_{diff}$ for SF-SF pairs compared to control samples for the stellar mass ratio, $\rm \Delta log(SFR)$, and $\rm \Delta log(sSFR)$. The horizontal solid lines, dotted lines represent the median value, the $1 \sigma$ confidential intervals in each bin, respectively. The top, bottom panels represent the metallicity indicator of D02, KD02, respectively.}
		\label{delta_mass_sfr_ssfr_dr7}
	\end{figure*}

	We also use the violin plots to analyze the parameters that might influence the distribution of $\rm \Delta log(O/H)_{diff}$, such as stellar mass ratio, SFR, and sSFR, as shown in Figure \ref{delta_mass_sfr_ssfr_dr7}. An additional requirement is included: the pairs and control samples should have similar SFR, sSFR (0.1 dex) for the left and middle panels, respectively. There is no significant difference between the distribution of pairs and control samples in each panel, which means that the SFR, sSFR, and stellar mass might not significantly change the $\rm \Delta log(O/H)_{diff}$ between control samples and pairs. A slight increase of $\rm \Delta log(O/H)_{diff}$ with an increase of $\rm \Delta log(SFR)$ can be found in the left two panels. For the right four panels, a slight negative trend can be found in the relationship between $\rm \Delta log(O/H)_{diff}$ and $\rm \Delta log(sSFR)$, and $\rm \Delta log(O/H)_{diff}$ and the stellar mass ratio. Overall, the SFR difference, the sSFR difference, and the stellar mass ratio are found not to influence the $\rm \Delta log(O/H)_{diff}$ between pairs and control samples.

\begin{figure*}
		
		\centering
		\begin{multicols}{2}
			\includegraphics[width=\columnwidth]{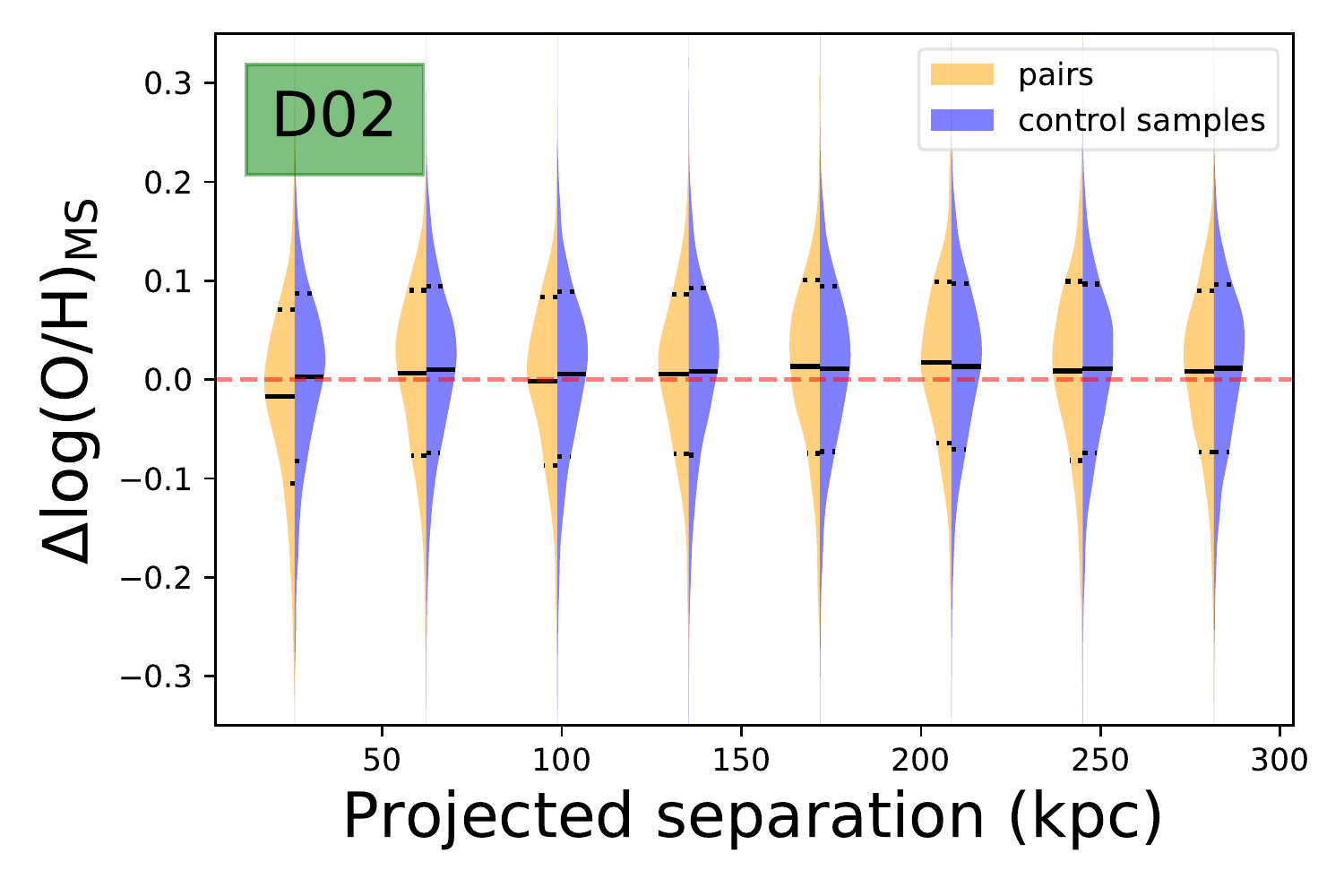}\par
			\includegraphics[width=\columnwidth]{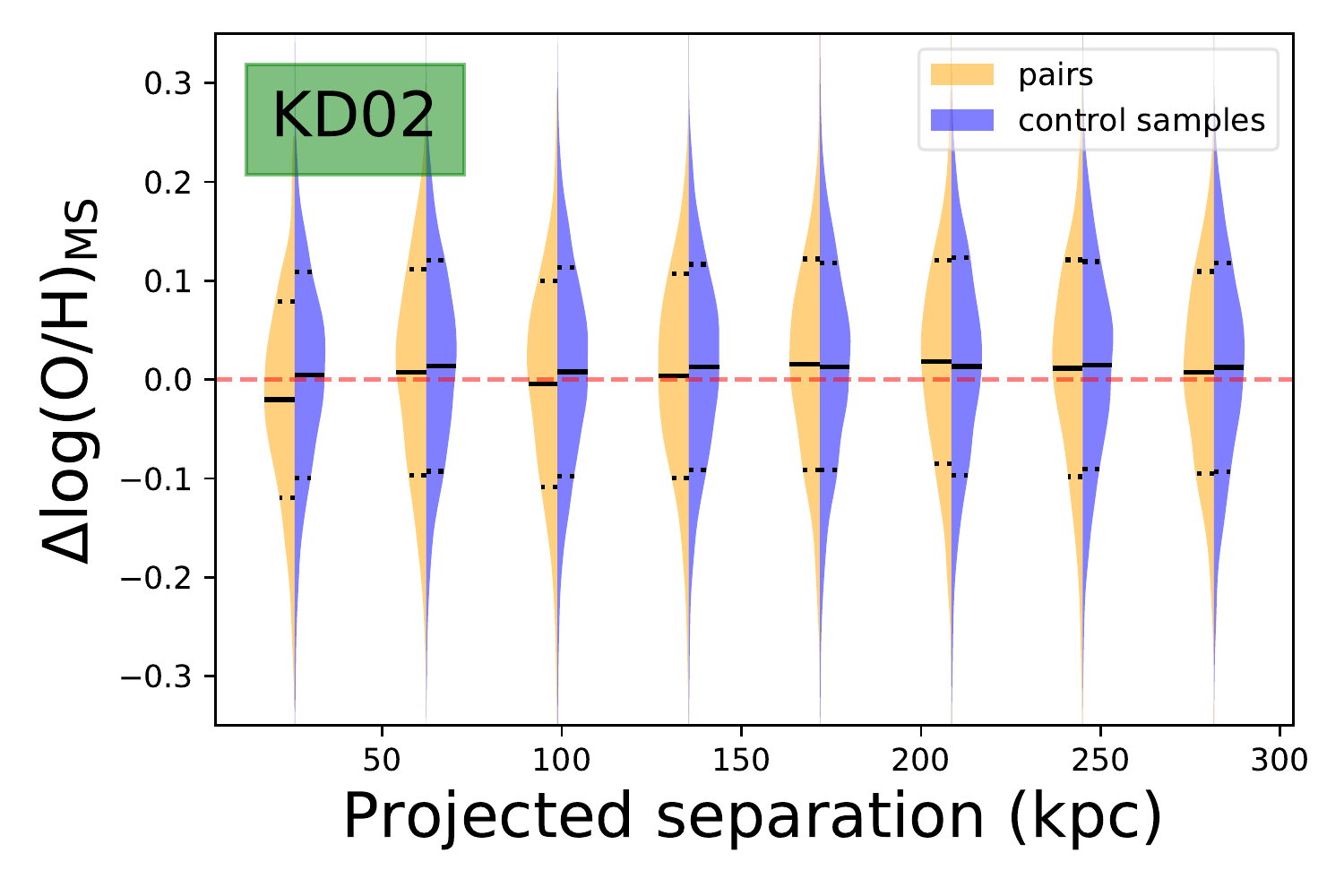}\par
		\end{multicols}
		\begin{multicols}{2}    
			\includegraphics[width=\columnwidth]{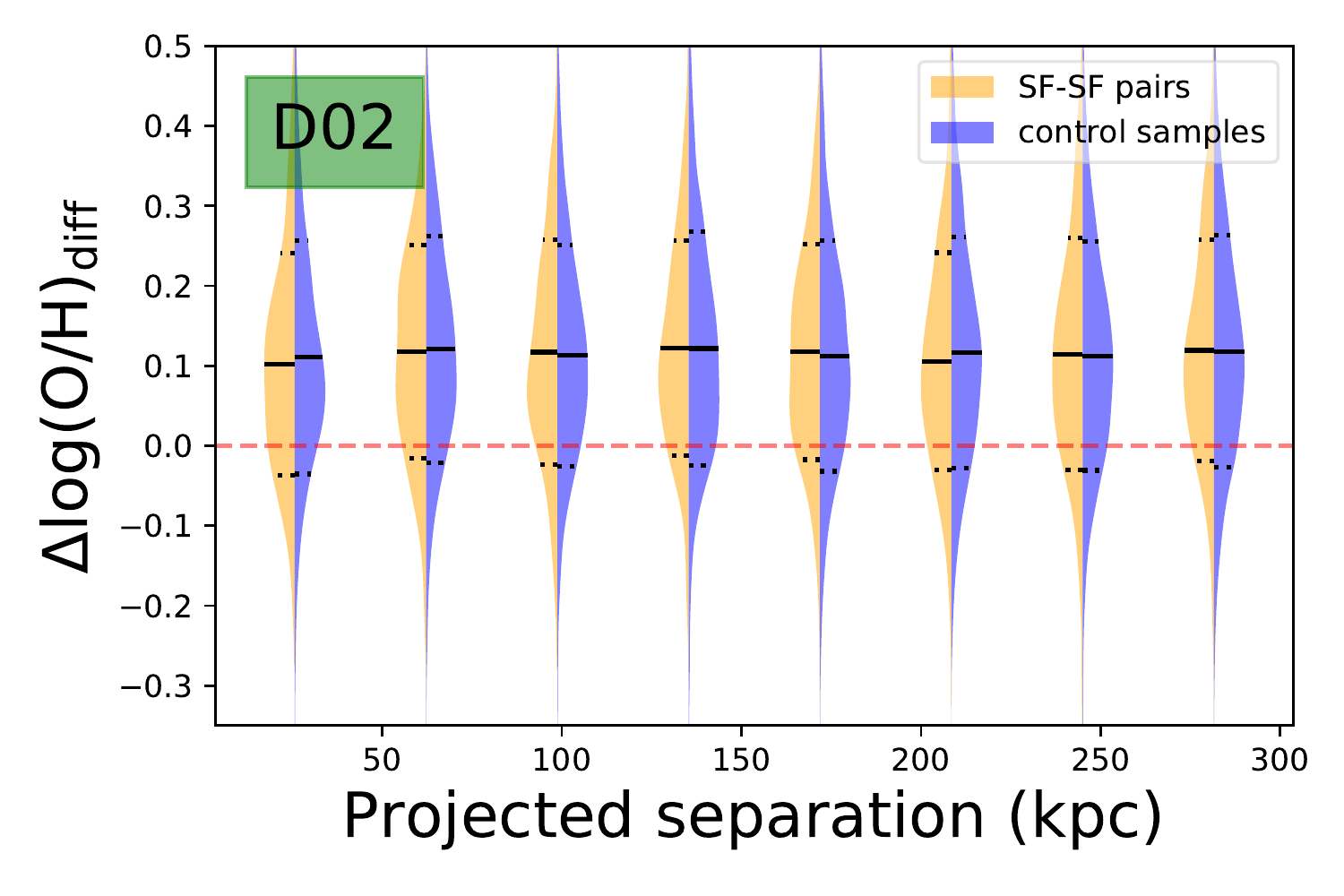}\par
			\includegraphics[width=\columnwidth]{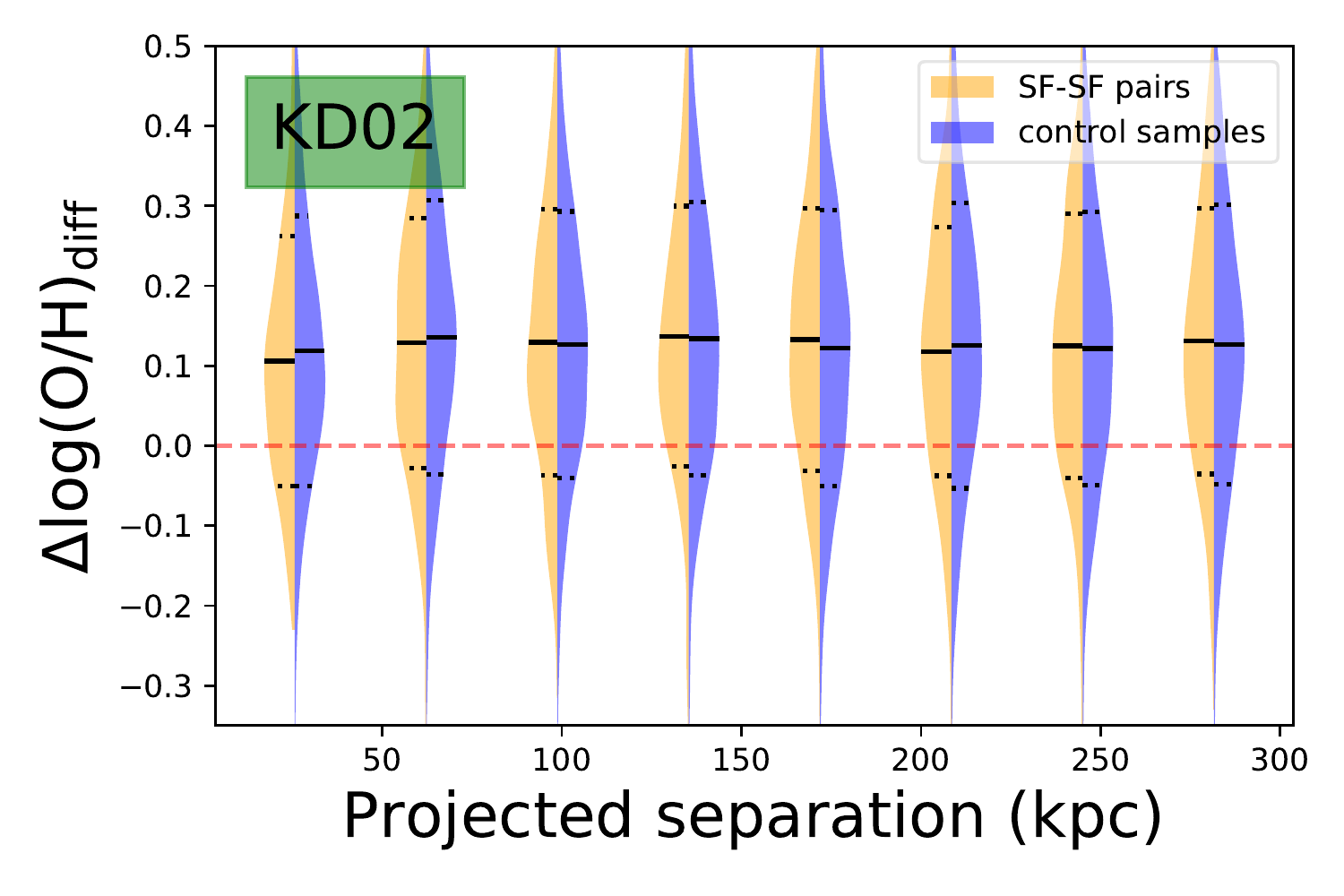}\par
		\end{multicols}    
    	\caption{Violin plots show the probability distribution of $\rm \Delta log(O/H)_{MS}$, $\rm \Delta log(O/H)_{diff}$ for both SF-SF pairs and SF-Passive pairs, SF-SF pairs, respectively, compared to control samples for the projected separation. In the violin plot, the white spot, the thick black bars, the thin black bars, the thin black lines in each bin represent the median value, the interquartile range, 1.5 $\times$ interquartile range, the distribution of y-axis data, respectively. The metallicity indicator in the left, the right panel are from D02, KD02, respectively. The red horizon line in each panel represents the y-axis data equals to zero.}
		\label{separation_dr7}
	\end{figure*}

	The trend of $\rm \Delta log(O/H)_{MS}$ and $\rm \Delta log(O/H)_{diff}$ as a function of the projected separation is further used to investigate whether the separation between pairs impacts the metallicity distribution. The closer the two galaxies, the larger area of sharing the CGM. We expect to see that the projected distance between pairs might influence the metallicity distribution. In the top panels of Figure \ref{separation_dr7}, the median value of $\rm \Delta log(O/H)_{MS}$ for pairs is smaller than that of control samples, when the projected distances are smaller than $\rm 150 \ kpc$. However, in the bottom panels, no evidence shows the difference of $\rm \Delta log(O/H)_{diff}$ between pairs and control samples. Compared to previous work (e.g.,
	\cite{2012MNRAS.426..549S}, \cite{2013MNRAS.435.3627E}, \cite{Bustamante2020}) that found metallicity dilution, no significant metallicity dilution is found in this work. This difference is caused by the different definition of control samples (section \ref{control sample} in detail). Because the more massive galaxies contain more metal, the median value of $\rm \Delta log(O/H)_{diff}$ is greater than zero. However, there is no obvious trend of change as a function of the projected distance.

	\section{Discussion}
	\subsection{Incompleteness}
	
	Due to the incompleteness, we might lose some pairs in our sample and thus obtain a biased result. The incompleteness of data is caused by the flux limit of the survey and the fiber collision effect \citep{2016MNRAS.461.2589P}. In this part, we will discuss the influence of incompleteness on the results.
	
\begin{figure}
	
	\centering
		\includegraphics[width=0.5\textwidth]{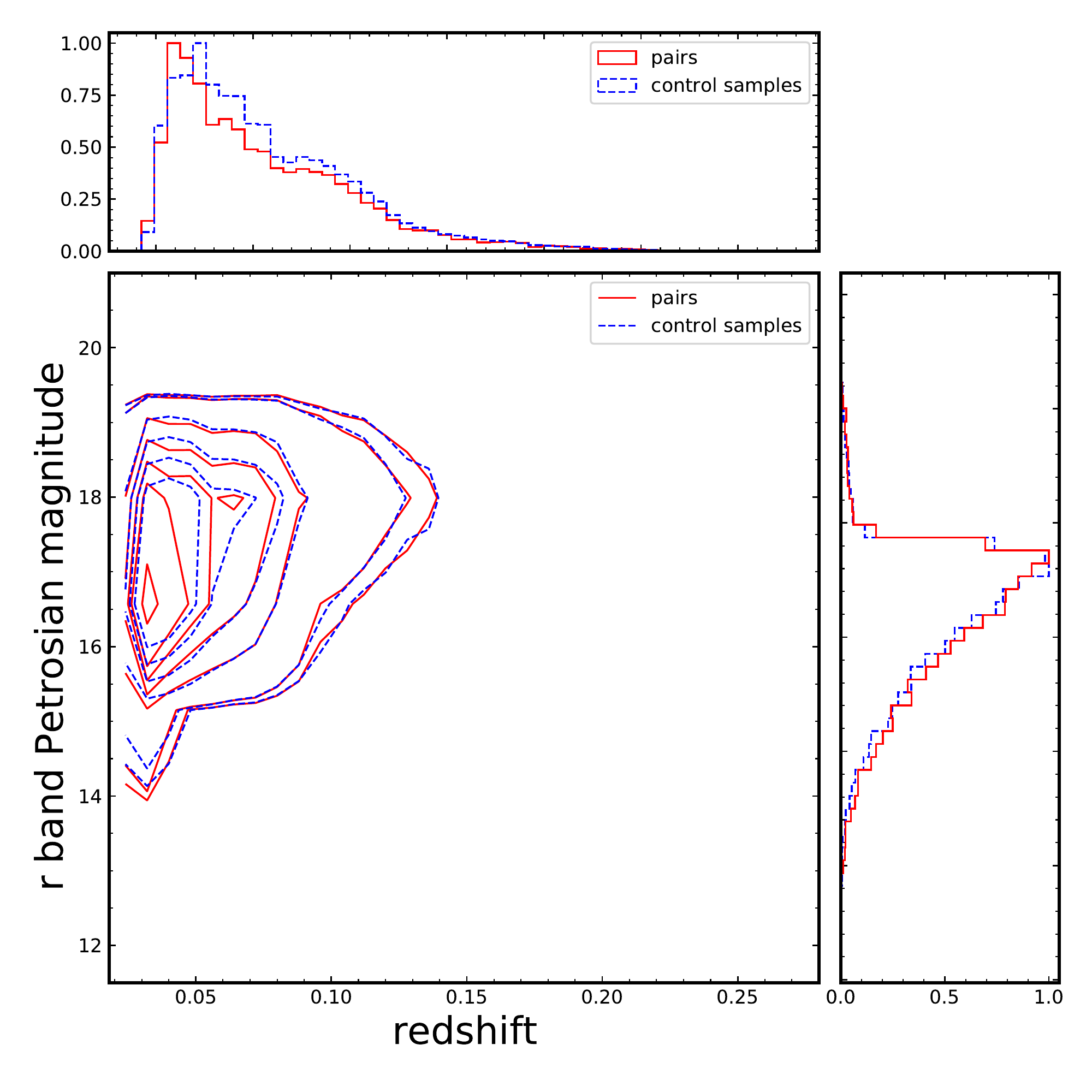}
	    \caption{The relationship between redshift and r band Petrosian magnitude. The horizontal and vertical histogram represents the distribution of redshift and the r band magnitude, respectively.}
	    \label{rband}
    \end{figure}

	Firstly, for the flux limit of the SDSS survey, we separately analyze the distribution of the Petrosian r band magnitude for pairs and control samples. In Figure \ref{rband}, the r band distribution of these two sets is quite similar. However, the distribution of redshift shows that there are more control samples lies between 0.05 and 0.2 than pairs. A weighting scheme in \S2 is introduced to reduce the difference. Even though, the overall distribution of Petrosian r band magnitude between pairs and control samples is similar because of the selection requirement. However, a small difference exist in redshift distribution between pairs and control sample. Thus, the incompleteness caused by the flux limit of the survey of control samples and matched pairs is similar.  
	
    Secondly, the existence of the physical separation of fibers will make some pair galaxies missed if they are located out of the fiber. The phenomenon that pairs appear in the photometric data but not in the spectroscopic data is called spectroscopic incompleteness. For SDSS, the fiber separation is $\rm 55 \ arcsec$ \citep{2003AJ....125.2276B,2016MNRAS.461.2589P}. 
    \cite{2008ApJ...685..235P} found that the ratio of the spectroscopic to photometric pairs at angular distance $\rm \ge 55 \ arcsec$ is 37.5\% of that at angular distance $\rm \le 55 \ arcsec$. Thus, for pairs whose the projected distances are less than $\rm 55 \ arcsec$, we use a weight scheme $w_{\theta} = 3.08$ to correct this selection effect \citep{Bustamante2020}.
	
	 As seen in Figure \ref{ratio}, there is no significant difference between the left (unweighted) and the right (weighted) panels. Table \ref{metal_diff} and \ref{offset} present a quantitative comparison. In this case, we find that the incompleteness introduced by fiber collision does not significantly influence the results.
	 
	 \subsection{The effects of different control samples}
	 \label{control sample}
\begin{figure*}
	
	\centering
	\begin{multicols}{2}
	    \includegraphics[width=\columnwidth]{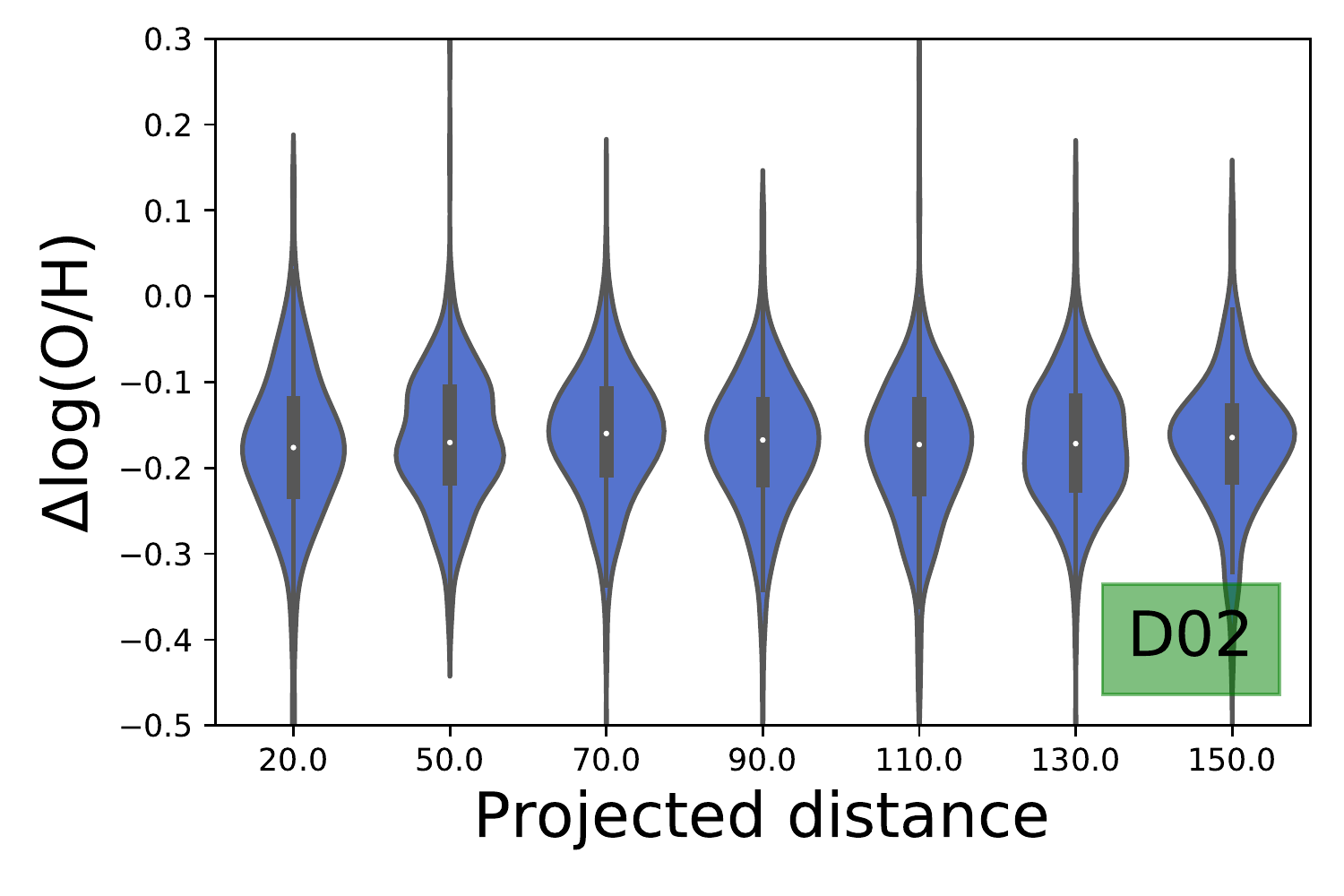}\par
		\includegraphics[width=\columnwidth]{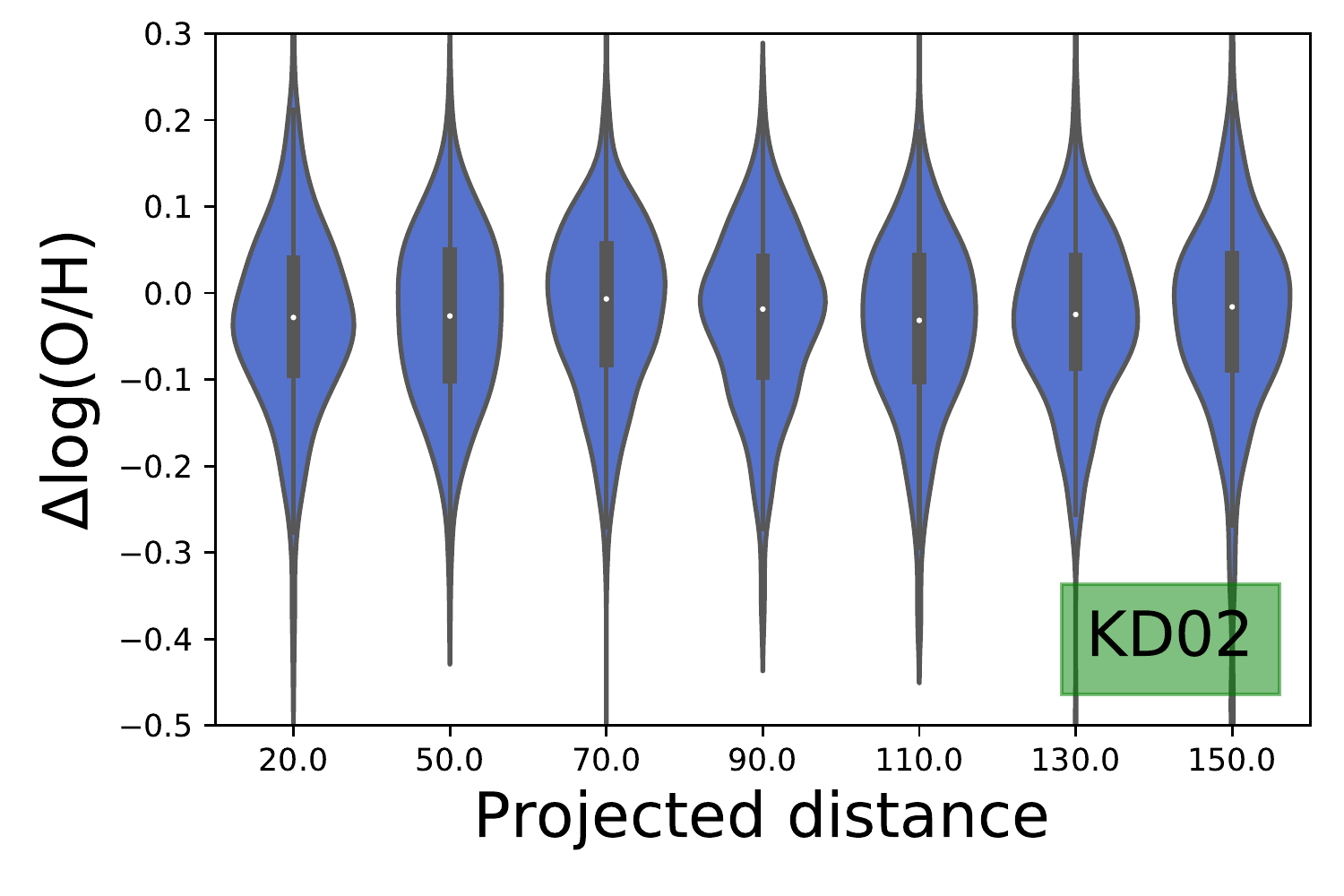}\par
	\end{multicols}
	 \begin{multicols}{2}
	    \includegraphics[width=\columnwidth]{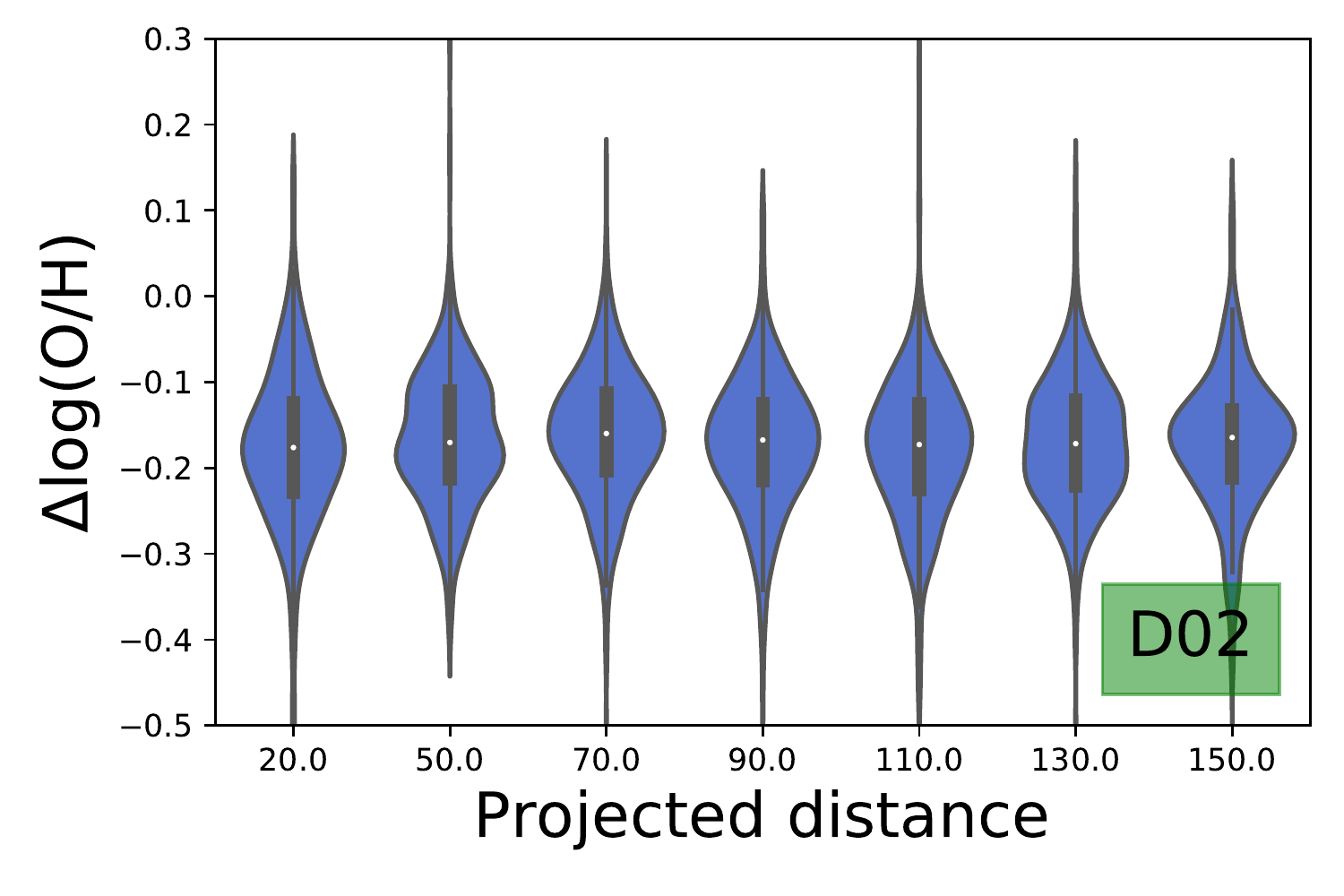}\par
		\includegraphics[width=\columnwidth]{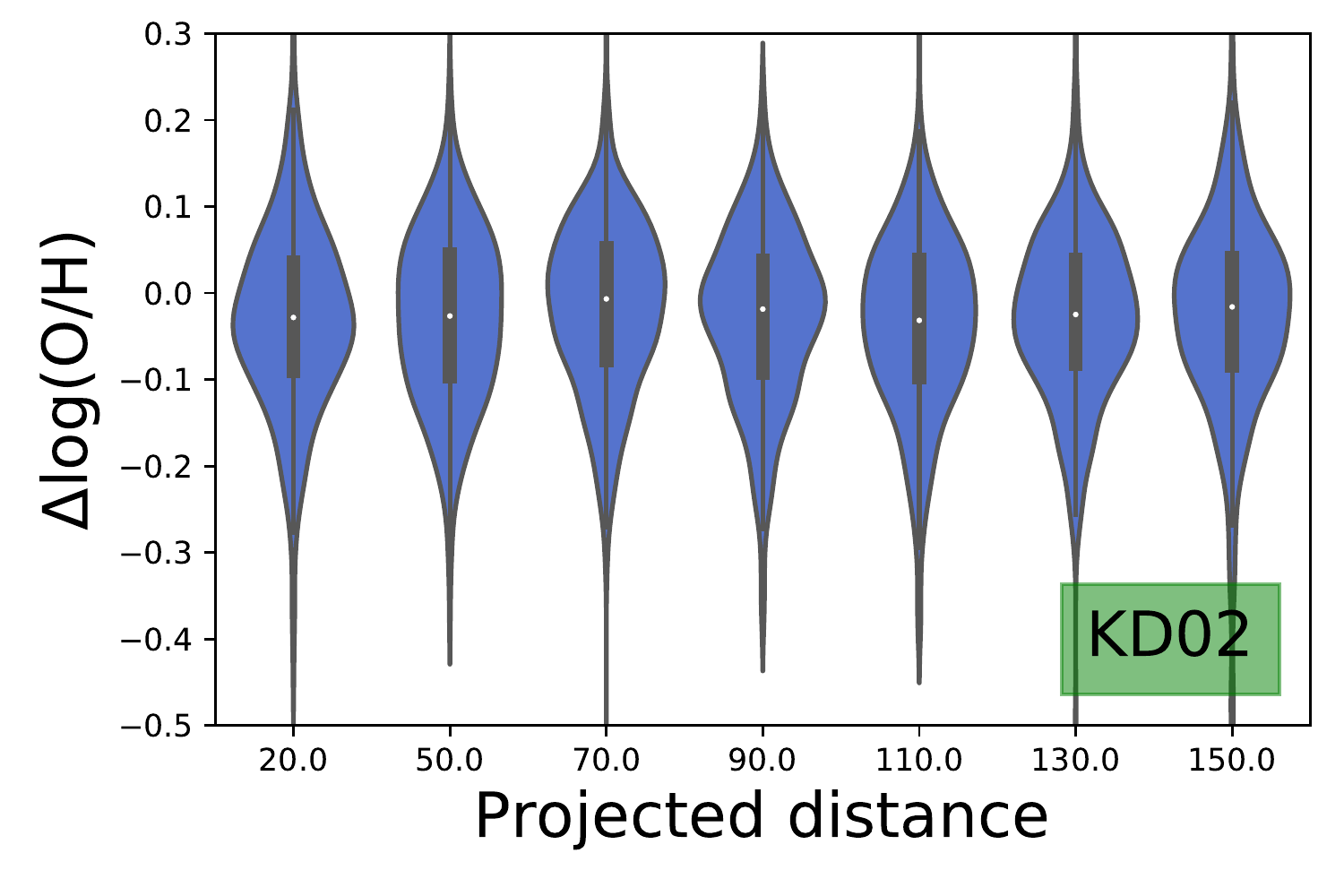}\par
	 \end{multicols}  
		
	    \caption{The relationship between $\rm \Delta log(O/H)$ and the projected distance. The $\rm \Delta log(O/H)$ in the top panels are calculated by using equation \ref{'equationmedian'}. The $\rm \Delta log(O/H)$ in the bottom panels are calculated by using the FMR from \protect\cite{Bustamante2020}.}
	    \label{fmr}
    \end{figure*}

	 For $\rm \Delta log(O/H)_{MS}$, the different definition of control samples will influence the final result. Here, we compare our result with that from \cite{2012MNRAS.426..549S}, which used the following equation:
	 
	  \begin{equation}
     \rm \Delta log(O/H)= 12+log(O/H)_{pair}-median(12+log(O/H)_{control}),
    \label{'equationmedian'}
    \end{equation}
    where $\rm median((12+log(O/H))_{control})$ is the median value of control samples for each pair galaxy. Because equation \ref{'equationmedian'} directly compares the metallicity between pairs and their matched control samples, we compare the metallicity of pairs with the median value of metallicity in each stellar mass bin of control samples. We find obvious metallicity dilution when the projected distance is smaller than 150 kpc, as shown in the top panels of Figure \ref{fmr}.
    
    We also plot the $\rm \Delta log(O/H)_{MS}$ using the function from \cite{Bustamante2020}, which use the fundamental metallicity relation (FMR). The FMR describes the relationship between the stellar mass, metallicity and SFR. In the bottom panels of the Figure \ref{fmr}, we find the metallicity dilution as a function of the projected distance too.

	Therefore, the definition of control samples typically affect the result of $\rm \Delta log(O/H)_{MS}$.

	\subsection{Physical Mechanism}
	
 No significant evidence is found in this paper that shows the sharing CGM will reduce the metallicity difference between pairs. This means that the sharing CGMs has a little measurable impact on regulating the metallicity of pairs. This agrees with the result from \cite{2016ApJ...822..107G} that the CGM metallicity has little effect on the metallicity in the galaxies. In this way, even with sharing CGMs that may mix and regulate the metal distribution in the CGMs between pairs, the change in the CGMs may not influence the metallicities within galaxies. In addition, as shown in \cite{2020MNRAS.499.2462P}, \cite{2021MNRAS.tmp..105W}, the metallicity distribution in the CGM is uneven: more metals located along the minor axis than the major axis of galaxy. Outflows that transport the metal-rich gas into the CGM, usually locate along the minor axis. Inflows can contain the metal-poor gas, usually locate along the major axis. As a result, the axis along which CGM is shared may fluctuate the metallicity dilution effects.

Compared to isolated galaxies, pairs have an additional way to affect the distribution of gas between each other. When galaxies begin to approach, they will first share the CGM and then have interaction with each other. Previous studies \citep{2008AJ....135.1877E, Bustamante2020} show that the interaction between pairs will dilute the gas and enhance the star formation. In this work, we expect to find the galaxies in pairs that have CGM sharing. So we redefine galaxy pairs to allow them to have a larger separation than that of pairs from previous works. Therefore, there exists an area where galaxies have CGM sharing but without interaction, which is shown in Table \ref{condi_diff}.

\begin{figure*}
		
		\centering
		\begin{multicols}{2}
			\includegraphics[width=\columnwidth]{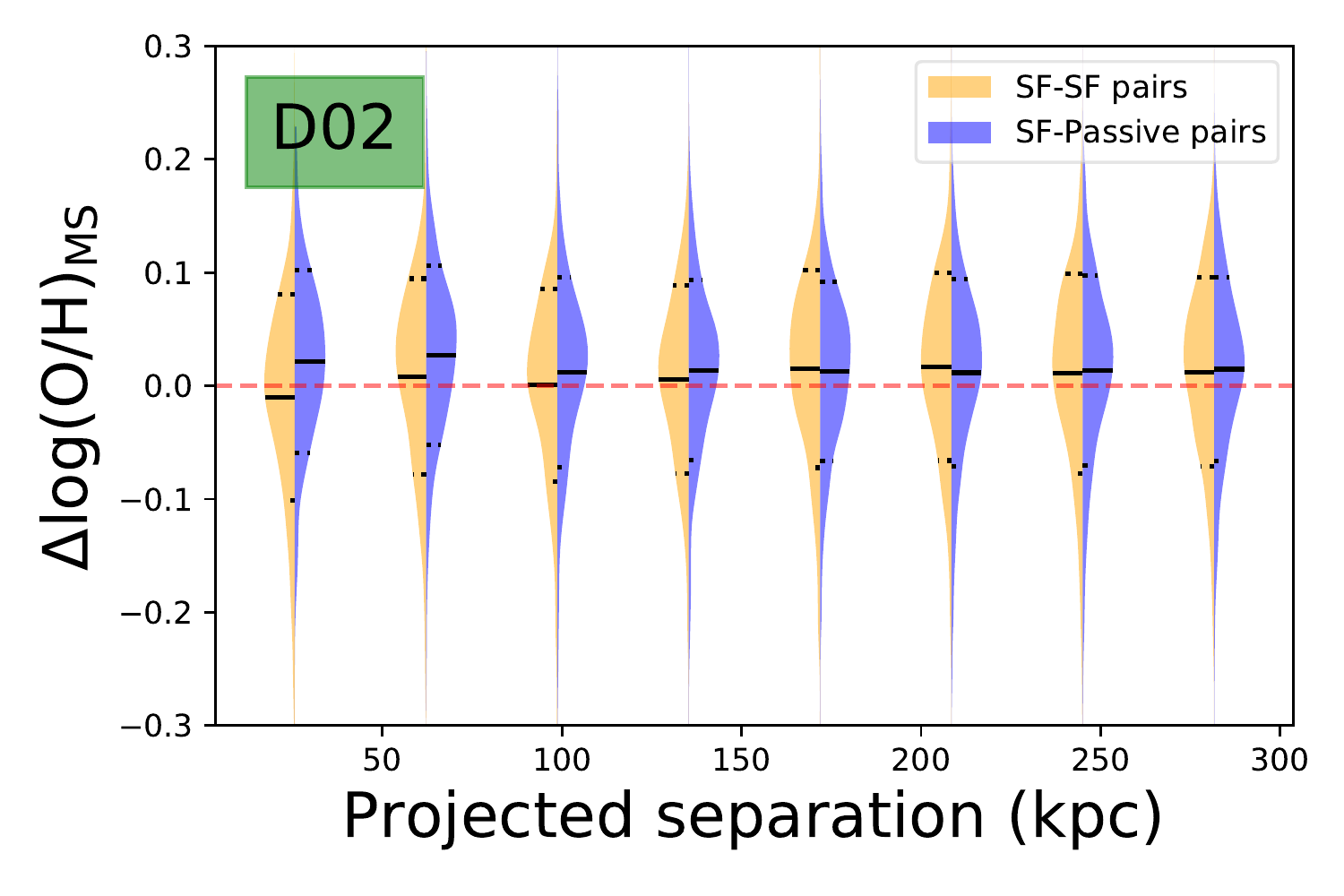}\par
			\includegraphics[width=\columnwidth]{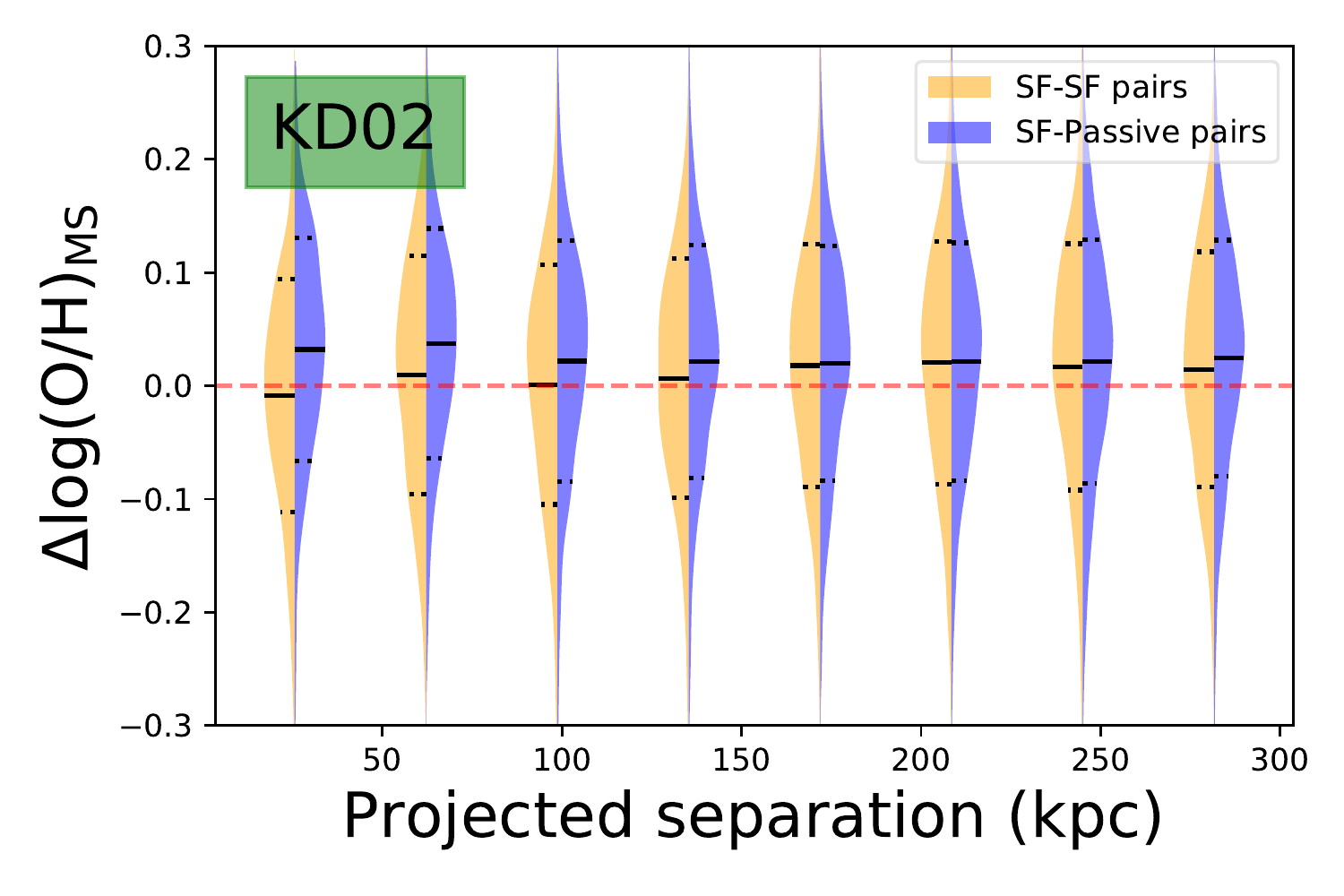}\par
		\end{multicols}
		
		%\begin{multicols}{2}    
		%	\includegraphics[width=\columnwidth]{./figure/sigma_sep_passive_median_do2}\par
		%	\includegraphics[width=\columnwidth]{./figure/sigma_sep_passive_median_kd02}\par
		%\end{multicols}    
		
    	\caption{Violin plots show the probability distribution of $\rm \Delta log(O/H)_{MS}$ for SF-SF pairs compared to SF-Passive pairs as a function of the projected separation. In the violin plot, the white spot, the thick black bars, the thin black bars, the thin black lines in each bin represent the median value, the interquartile range, 1.5 $\times$ interquartile range, the distribution of y-axis data, respectively. The metallicity indicator in the left, the right panel are from D02, KD02, respectively. The red horizon line in each panel represents the y-axis data equals to zero. The orange, blue color in the picture represent the SF-SF pairs and SF-Passive pairs respectively. The $\rm \Delta log(O/H)_{MS}$ equal to the metallicity offset from the "best-fitted" stellar mass-metallicity relationship.}
		\label{'sf-passive'}
	\end{figure*}

In Figure \ref{'sf-passive'}, when the projected distance $\leq \rm 150 kpc$, the $\rm \Delta log(O/H)_{MS}$ of SF-SF pairs are smaller than that of SF-Passive pairs in each separation bin, which shows that the interaction triggers the metal-poor gas from the sharing CGM to dilute metals within galaxies. When the projected distance $\geq \rm 150 kpc$, there is no significant difference in $\rm \Delta log(O/H)_{MS}$ between SF-SF pairs and SF-Passive pairs, which means that only the sharing CGM itself, without the galaxy interaction, don't have the possibility to affect the metal distribution between pairs. Because the majority of the SF-Passive pairs are minor mergers, we add a stellar mass ratio cut on SF-SF pairs. The result doesn't change, so we think it is valid in the minor merger case. This again proves that the sharing of CGM without interactions may not impact the central metallicity of galaxies.
  
	\section{Conclusions}
	
	In this work, we select star-forming galaxies with strong emission lines and explore the MZR of pair galaxies using single fiber data from SDSS DR7. To study the effect of CGM on regulating the star formation and the metal re-distribution process, pairs are defined to have separations of less than 300 kpc. A weighting scheme is introduced to correct the select effect from flux limit and fiber collision. The main results are shown below:
	
	(1) The MZR trend is similar to \cite{2004ApJ...613..898T}: the monotonic correlation between stellar mass-metallicity and the trend becomes flat when stellar mass is higher than $ 10^{10.5} \  M_{\odot}$.
	
	(2) Compared to control samples, there is no significant difference in the distribution of metallicity offset from the best-fitted stellar mass-metallicity relationship of galaxies in SF-SF pairs ($\rm \Delta log(O/H)_{MS}$), metallicity difference between two members in pairs ($\rm \Delta log(O/H)_{diff}$, both SF-SF and SF-Passive pairs), which means that the CGM will not decrease the metallicity difference within $3''$ region of pairs. 
	
	(3) The parameters such as the stellar mass ratio, SFR, and sSFR are found not to impact $\rm \Delta log(O/H)_{MS}$ and $\rm \Delta log(O/H)_{diff}$ between pairs.
	
	(4) Comparing the $\rm \Delta log(O/H)_{MS}$ of SF-SF pairs with that of SF-Passive pairs, we find that only with the assistance of galaxy interaction, the sharing CGM can trigger the metal-poor gas fall into the galaxy center. Because most of the SF-Passive pairs are minor mergers, this result is valid in the minor merger case.

In conclusion, the CGM sharing should not be a major factor that shapes the metal evolution of galaxies. The gas recycling and the uneven distributions of metals in CGMs may fluctuate the effects of sharing CGMs in regulating metallicity of galaxies.

	\section*{Acknowledgements}
	We thank the referee for a detailed report that significantly improve the presentation of our work.
	S.Z. and Y.S. acknowledge the support from the National Key R\&D Program of China (No. 2017YFA0402704, No. 2018YFA0404502), the National Natural Science Foundation of China (NSFC grants 11825302, 11733002 and 11773013). Y.S. thanks the support from the Tencent Foundation through the XPLORER PRIZE. 
	
	\section*{Data availability}
    This paper makes use of the MPA-JHU DR7 data, which is available at \url{https://wwwmpa.mpa-garching.mpg.de/SDSS/DR7}.

	%%%%%%%%%%%%%%%%%%%%%%%%%%%%%%%%%%%%%%%%%%%%%%%%%%
	
	
	
	%\bibliographystyle{mnras}

	
	\bibliographystyle{mnras}
	\bibliography{1}

	%%%%%%%%%%%%%%%%%%%%%%%%%%%%%%%%%%%%%%%%%%%%%%%%%%
	
	%%%%%%%%%%%%%%%%% APPENDICES %%%%%%%%%%%%%%%%%%%%%

	%%%%%%%%%%%%%%%%%%%%%%%%%%%%%%%%%%%%%%%%%%%%%%%%%%

	% Don't change these lines
	\bsp    % typesetting comment
	\label{lastpage}
\end{document}